\documentclass[useAMS,fleqn,usenatbib]{mnras}

\usepackage{newtxtext,newtxmath}

\usepackage[T1]{fontenc}

\DeclareRobustCommand{\VAN}[3]{#2}
\let\VANthebibliography\thebibliography
\def\thebibliography{\DeclareRobustCommand{\VAN}[3]{##3}\VANthebibliography}

\usepackage{graphicx}	
\usepackage{amsmath}	
\usepackage{ulem}
\usepackage{cancel}
\usepackage{xcolor}
\usepackage{grffile}
\usepackage{url}
\usepackage{comment}
\usepackage{natbib}
\usepackage{enumerate}
\usepackage{verbatim}
\usepackage{float}
\usepackage{placeins}
\usepackage{soul}
\usepackage[colorinlistoftodos]{todonotes}

\newcommand\sect{Section}

\newcommand\eqn{Eq.}
\newcommand\eqns{Eqs.}

\newcommand\wlmass{M_{\rm{WL}}}
\newcommand\truemass{M_{\rm{T}}}
\newcommand\bias{b_{\rm{WL}}}
\newcommand\mth{M_{200}}
\newcommand\mfh{M_{500}}
\newcommand\zlens{z_{\mathrm{l}}}
\newcommand\zsource{z_{\mathrm{s}}}
\newcommand\esource{\epsilon_{\mathrm{s}}}
\newcommand{\appropto}{\mathrel{\vcenter{
  \offinterlineskip\halign{\hfil$##$\cr
    \propto\cr\noalign{\kern2pt}\sim\cr\noalign{\kern-2pt}}}}}

\newcommand {\apgt} {\ {\raise-.5ex\hbox{$\buildrel>\over\sim$}}\ }
\newcommand {\aplt} {\ {\raise-.5ex\hbox{$\buildrel<\over\sim$}}\ }

\defcitealias{2011ApJ...740...25B}{BK11}

\title[Weak lensing mass modeling bias]{Weak lensing mass modeling bias and the impact of miscentring}  

\author[Sommer et al.]{
Martin W. Sommer$^{1}$\thanks{E-mail: mnord@astro.uni-bonn.de (MWS)},
Tim Schrabback$^{1}$,
Douglas E. Applegate$^{1,2}$,
Stefan Hilbert$^{3}$,
\newauthor 
Behzad Ansarinejad$^{4}$,
Benjamin Floyd$^{5}$,
Sebastian Grandis$^{3}$
\\
\\
$^{1}$Argelander-Institut f\"{u}r Astronomie, Auf dem H\"ugel 71, D-53121 Bonn, Germany \\
$^{2}$Kavli Institute for Cosmological Physics, University of Chicago, 5640 South Ellis Avenue, Chicago, IL 60637, USA \\
$^{3}$Faculty of Physics, Ludwig-Maximilians-Universit\"{a}t, Scheinerstr.
1, 81679, Munich, Germany \\
$^{4}$School of Physics, University of Melbourne, Parkville, VIC 3010, Australia \\
$^{5}$Department of Physics and Astronomy, University of Missouri$-$Kansas City, 5110 Rockhill Road, Kansas City, MO 64110, USA \\
}

\date{May 18, 2021}

\pubyear{2021}

\begin{document}
\label{firstpage}
\pagerange{\pageref{firstpage}--\pageref{lastpage}}
\maketitle

\begin{abstract}
Parametric modeling of galaxy cluster density profiles from weak lensing observations leads to a mass bias, whose detailed understanding is critical in deriving accurate mass-observable relations for constraining cosmological models. Drawing from existing methods, we develop a robust framework for calculating this mass bias in one-parameter fits to simulations of dark matter halos. We show that our approach has the advantage of being independent of the absolute noise level, so that only the number of halos in a given simulation and the representativeness of the simulated halos for real clusters limit the accuracy of the bias estimation. While we model the bias as a log-normal distribution and the halos with a Navarro-Frenk-White profile, our method can be generalized to any bias distribution and parametric model of the radial mass distribution. We find that the log-normal assumption is not strictly valid in the presence of miscentring of halos. We investigate the use of cluster centers derived from weak lensing in the context of mass bias, and tentatively find that such centroids can yield sensible mass estimates if the convergence peak has a signal-to-noise ratio approximately greater than four. 
In this context we also find that the standard approach to estimating the positional uncertainty of weak lensing mass peaks using bootstrapping severely underestimates the true positional uncertainty for peaks with low signal-to-noise ratios.
Though we determine the mass and redshift dependence of the bias distribution for a few experimental setups, our focus remains providing a general approach to computing such distributions. 
\end{abstract}



\section{Introduction}
\label{sec:introduction}

The abundance of clusters of galaxies at different epochs is highly sensitive to the geometry of the universe and to the integrated growth rate of primordial density fluctuations (e.g.~\citealt{2001ApJ...553..545H}). As a consequence, number counts of galaxy clusters as a function of mass and redshift are a powerful tool for investigating the dark energy equation of state and other parameters of the standard cosmological model \citep[for a review see, e.g.,][]{2011ARA&A..49..409A}, and potentially also in testing for deviations from the predictions of general relativity on the scale of the universe as a whole (e.g. ~\citealt{2010MNRAS.406.1796R}).

The absolute calibration of mass-observable relations is an important factor in deriving accurate cosmological constraints based on the galaxy cluster abundance. 
Within the framework of general relativity, clusters of galaxies deflect light astigmatically, giving rise to distortions in the images of background galaxies. While this effect (weak lensing, henceforth WL) currently provides the most direct method of calibrating cluster masses, there are various sources of bias that must be carefully accounted for. Currently, uncertainties in such biases contribute significantly to the overall systematic error budget of mass-observable relations based on WL measurements (e.g. \citealt{2014MNRAS.439...48A, 2014MNRAS.440.2077M,2016A&A...594A..24P,2018MNRAS.474.2635S, 2019MNRAS.483.2871D,2019ApJ...878...55B,2020arXiv200907591S,2019MNRAS.482.1352M}). It is thus crucial to obtain a better understanding of the mass bias and its dependencies. The goal of this work is to improve our understanding of the mass bias arising due to fitting parametric radial density models to shear profiles derived from WL measurements.

\cite{2019MNRAS.488.2041G} estimated the expected contributions to the systematic error budget of the absolute mass calibration in the planned Euclid\footnote{http://sci.esa.int/euclid/} \citep{2011arXiv1110.3193L} and Rubin Observatory Legacy Survey of Space and Time\footnote{https://www.lsst.org/} \citep[LSST,~][]{2019ApJ...873..111I} surveys, and predicted uncertainties on the order of one per cent from sources not directly related to mass bias (accuracy of shape measurements, mis-estimation of lensing efficiency, and uncertainties in the estimation of contamination from cluster members). While the mass bias modelling accuracy and uncertainty ideally need to match this 1\% level of systematic uncertainty in order to not degrade the constraining power of future surveys substantially, direct current constraints vary in the range of 3-5\% \citep[e.g.][]{2019MNRAS.483.2871D}.

For a perfectly centered shear profile, the mass bias can be viewed as coming from three distinct contributions: the use of a parametric model, the triaxial mass distribution of galaxy clusters (or the presence of complex substructures, such as in merging systems), and large-scale structure along the line of sight (where the latter can be separated into correlated and uncorrelated contributions). In addition, shear profiles are of course never perfectly centered, giving rise to a fourth contribution related to miscentring. The latter can, as we shall discuss, be considered as a separate problem. 

Using parametric models of mass density as a function of radius to estimate masses of galaxy clusters invariably leads to bias even for an imagined perfectly spherical system (unless a model can be found that perfectly matches all such systems). In general, the level of bias will critically depend upon the radial range used for the mass analysis, as parametric models of galaxy clusters have been found to agree to different degrees with observations in different radial ranges. 

The triaxial shapes of cold dark matter (CDM) halos have been found to bias spherically symmetric model fits to profiles of tangential reduced shear profiles \citep{2001A&A...378..748K,2004MNRAS.350.1038C,2005ApJ...632..841O,2007MNRAS.380..149C,2010A&A...514A..93M}, with systematic offsets of up to $50\%$ in individual mass estimates depending on the geometry of the system. For halos elongated along the line of sight, masses tend to be overestimated, while masses tend to be underestimated in halos with major axes approximately perpendicular to the line of sight. 

Large-scale structure (henceforth LSS) along the line of sight can be subdivided into correlated and uncorrelated contributions, although the distinction is not straightforward. While uncorrelated LSS can add scatter to determined masses, it is not expected to significantly bias WL masses on average \citep{2001A&A...370..743H,2011MNRAS.412.2095H}. Systematic errors due to uncorrelated LSS can thus be decoupled from the determination of a weak lensing mass bias distribution, assuming that it can be reliably separated from the correlated LSS. 

The projection of LSS and the effects of triaxiality are not independent. Neighboring halos are generally connected by filaments, and the direction of the major axis of a halo is correlated with the directions to massive neighbors (e.g., \citealt{2009ApJ...706..747Z}, and references therein). Such alignments persist out to radii of approximately 100 $h^{-1}$ Mpc from the cluster center \citep{2002A&A...395....1F,2005ApJ...618....1H}, suggesting an optimal integration length of $\sim$200 $h^{-1}$ Mpc ($\pm 100$ $h^{-1}$ Mpc with the halo at zero) for separating correlated and uncorrelated LSS in cosmological simulations. 
For halos with mass $\mfh > 1.5 \times 10^{14} h^{-1} M_{\odot}$
\footnote{We define $M_{\Delta}$ (generally with $\Delta \in \{ 200,500 \} $) to be synonymous with $M_{\Delta , c}$, the mass enclosed within the corresponding radius $r_{\Delta , c}$ such that the average mass density within this radius is equal to $\Delta$ times the critical density of the Universe at the redshift of the halo, as defined in \sect~\ref{sec:method:nfw}.}, \citet[henceforth~\citetalias{2011ApJ...740...25B}]{2011ApJ...740...25B} found the WL mass bias distribution to be stable for integration lengths in a range of approximately 30-200 $h^{-1}$ Mpc comoving.
Here, we account only for correlated LSS, deriving mock WL data from simulations using similar integration lengths.  

At large radii, correlated matter around the cluster (correlated halos) contributes to the lensing profile (the so-called two-halo term, e.g. \citealt{2000MNRAS.318..203S}, \citealt{2005MNRAS.362.1451M}). In this work, we limit the outer radius so as to make this term negligible. 

While azimuthally symmetric radial models are often used in weak lensing analyses \citep[e.g][]{2016MNRAS.457.1522A,2019MNRAS.483.2871D}, one must carefully account for how the center coordinate is chosen. This choice can be made in different ways: first, the center may be modeled directly in conjunction with the shear profile. Second, it may be derived from the peak in a different observable such as Compton-Y (Sunyaev-Zeldovich effect, also SZE, \citealt{1970CoASP...2...66S, 1980ARA&A..18..537S}), X-ray emissivity or some observable derived from the distribution of cluster galaxies. Third, one may use the peak in the weak lensing convergence (derived up to a constant from the reduced shear). Fourth, for very massive clusters a strong-lensing derived center may be used for the WL analysis, as done by, e.g., \cite{2018A&A...610A..85S}. Of course it is also possible to combine these methods. For example, the X-ray emissivity peak, combined with its uncertainty, may be used as a prior when fitting for the centroid.  

We define the term \textit{miscentring} as the absolute projected offset between the employed center  and the true cluster center, where the latter is defined as the position of the most bound particle in the simulation, corresponding approximately to the bottom of the gravitational well. 

There are essentially two ways to account for miscentring. The miscentring can be modeled as part of the mass bias (e.g. \citealt{2019ApJ...878...55B}), or it can be treated separately in forward modeling of masses from the shear profiles. Either way, a miscentring distribution, corresponding to the chosen center proxy, must be assumed. A derivation of the latter is not trivial. In particular, such a distribution would be expected to be anisotropic; for a merger elongated in the plane of the sky there would be preferred directions for miscentring; the X-ray center, for example, would typically be in the direction of one of the main gas clumps.

We focus on two main issues related to miscentring. First, we analyze how a miscentring distribution derived from Compton-Y images, similar to the ones produced by the South Pole Telescope (SPT, \citealt{2011PASP..123..568C}), impacts the weak lensing mass bias. Second, we derive centers as the peaks in the signal-to-noise ratio maps of the reconstructed convergence field, and compare their performance to the analysis using SZE centers. In this context, we also investigate the robustness of a common method for deriving positional uncertainties in convergence-derived center positions, based on bootstrapping the source galaxy sample. 

In recent years, significant efforts have been made towards quantifying the weak lensing mass bias distribution. We summarize the most important results of these works in the following paragraphs.

Using n-body simulations of dark matter only (DMO), \citetalias{2011ApJ...740...25B} first studied the scatter and bias
in WL mass measuremenes from Navarro-Frenk-White (NFW, \citealt{1997ApJ...490..493N}) reduced shear profile fits.
They took contributions from
matter located within the halo virial radius as well as correlated and
uncorrelated LSS into account, and generally found bias levels of five to
ten percent. The analysis showed that fitting beyond the virial radius biases masses low due to deviations from the NFW model at large radii, e.g. from neighboring halos.  
\citetalias{2011ApJ...740...25B} also considered miscentring to some extent, and found that halo centering errors can introduce negative mass bias at around 5\%. For both correlated and non-correlated LSS, the authors found a non-negligible contribution to the scatter, but none to the mean bias.

\cite{2011MNRAS.414.1851O} used the DMO simulations of \cite{2009ApJ...701..945S} to investigate mass biases. Including scales out to many times the virial radius, the authors fitted the simulated reduced shear profiles with a truncated NFW model (avoiding a divergence of the mass) plus a two-halo component (e.g. \citealt{2007ApJ...656...27J}), which accounts for the impact of neighbouring halos. In two-parameter fits (mass and concentration), the results were consistent with those of \citetalias{2011ApJ...740...25B} in terms of mass, with the concentration parameter typically being overestimated. \cite{2012MNRAS.421.1073B}, using the Millennium simulation \citep{2005Natur.435..629S} to study mass biases, found results consistent with \cite{2011ApJ...740...25B}. 

While n-body simulations modeling only dark matter are sufficient for many purposes, the focus has been gradually shifted to hydrodynamic simulations to account for the baryonic component. The inclusion of baryonic effects leads to high-mass clusters appearing more spherical and to higher concentrations on average \citep{2010MNRAS.405.2161D,2013MNRAS.429.3316B}. Simulations not including AGN feedback suffer from overcooling. In the absence of heating in the central region, heat dissipation is overly efficient, leading to overestimates in stellar fractions (e.g. \citealt{2011ASL.....4..204B}).

\cite{2017MNRAS.465.3361H} used both hydrodynamic and n-body simulations to quantify
how the inclusion of baryons affects
the WL mass bias, finding very similar results from both types of simulation. In particular, a mass bias consistent with \cite{2011ApJ...740...25B} was found at low masses, with the bias essentially vanishing at the highest cluster masses.

\cite{2018MNRAS.479..890L} used cosmoOWLS simulations \citep{2014MNRAS.441.1270L}, including baryons, and let both the mass and concentration vary freely, since the concentration$-$mass relation is sensitive to baryons. The results were consistent with those obtained by \cite{2017MNRAS.465.3361H} in the sense that differences between DMO and baryonic simulations, in terms of weak lensing mass bias, are very small. Weak lensing masses were found to be underestimated by around $10\%$ for low-mass systems ($M_{200} \simeq 2 \times 10^{14} M_{\odot}$), with the bias decreasing for higher mass clusters and consistent with no bias for the most massive systems, fully consistent with previous studies. Importantly, \citeauthor{2018MNRAS.479..890L} found some dependence on the absolute level of shape noise, which makes modeling the mass bias quite complicated in practice. However, there was no dependence on the noise level at the highest masses studied.

In summary, previous publications studying the mass bias have concluded that WL masses are underestimated by $0-10\%$, with the bias generally decreasing with increasing mass, both using DMO simulations  \citep{2011ApJ...740...25B,2011MNRAS.414.1851O,2012MNRAS.421.1073B} and simulations including baryons \citep{2017MNRAS.465.3361H,2018MNRAS.479..890L}.

In the coming years we expect that large-volume simulations including baryons will yield WL mass bias constraints with percent-level accuracy and enough halos to provide the matching precision. We however limit the scope of this paper to DMO simulations for two reasons. First, we seek to identify robust and general methods not specific to any particular simulation.
Second, the bias is sensitive to a number of factors (including the radial range of measured shear as well as the mass and the redshift of the cluster) that still dominate over the small differences in bias between baryonic and DMO simulations found by \cite{2018MNRAS.479..890L} and \cite{2017MNRAS.465.3361H}.

For the future, it will be crucial to use simulations that include baryon physics for the determination of mass bias, as these will allow us to simultaneously derive all relevant observables, including shear profiles and the WL center proxies as derived from Compton-Y and X-ray luminosity maps. Simulations including baryons will therefore play a crucial role in deriving spatial distributions for quantifying biases related specifically to miscentring. This will be the subject of a future work (Sommer et al., \textit{in prep.}). 

In this paper we develop, based on previous findings from the literature, a robust scheme for determining the weak lensing mass bias over a broad range of cluster masses and at any redshift accessible by simulations, such that the determined mass bias is not dependent upon the absolute noise level. We find that this is possible, under certain conditions, for one-parameter fits, fitting only to mass, and scaling for the concentration parameter (see \sect~\ref{sec:method:nfw}) from relations known from the literature. We explore some commonly used miscentring distributions, and contrast these with the ones resulting from using the peak of the lensing convergence as a center proxy. We also investigate whether the miscentring can be included in the mass bias in a robust way. In this context, we also touch upon the problem of deriving uncertainties on the position of the convergence peak, and find that a traditional bootstrapping technique severely underestimates this uncertainty. 

The weak lensing mass bias is a complicated function of many factors such as mass, redshift, the concentration$-$mass relation and the radial range of the analysis. To model it in a general way is thus difficult. Instead, the bias is better modeled individually for each cluster of galaxies studied, or for sub-samples of similar targets within a survey. However, it is still useful to identify how these factors influence the mass bias. First, this allows us to adapt the data analysis in a way that minimizes systematic uncertainties in the bias estimates. Second, it yields insight into what factors need to be constrained more tightly (for example, the miscentring distribution) to reduce this systematic uncertainty further. 

The present work is structured as follows. We describe the simulations, the mass selection and the parametric models used in \sect~\ref{sec:method}. Here we also describe the Bayesian framework used for deriving the weak lensing bias as a log-normal distribution in the presence of noise. In \sect~\ref{sec:results} we show under what circumstances the bias is noise independent and investigate how the bias depends upon various factors such as mass, redshift and the radial range of fitting. We derive a set of lensing-based miscentring distributions and compare these to corresponding distributions from SZE and X-ray data. We discuss the implications of our results for WL observations in \sect~\ref{sec:discussion}. We offer our conclusions and summarize our findings in \sect~\ref{sec:conclusion}. 

For the majority of our results we use the cosmological model corresponding to the MCMC simulations (described in \sect~\ref{sec:method:sim}), namely a flat $\Lambda$CDM cosmology with $h = 0.73$, $\Omega_{m}=0.25$ and $\Omega_{\Lambda}=0.75$. Where other cosmological models are used (because other simulations are based on different cosmologies), we state this explicitly.

\section{Method}
\label{sec:method}

\subsection{Weak lensing formalism}
\label{sec:meth:wl}

Gravitational lensing  by a foreground mass (the ``lens'') at redshift $\zlens$ introduces a distortion in the images of a background (``source'') galaxy at redshift $\zsource$. The convergence $\kappa(\btheta)=\Sigma(\btheta) / \Sigma_{\mathrm{crit}}$ at position $\btheta$ is the ratio of the surface mass density $\Sigma(\btheta)$ and the critical density
\begin{equation}
\Sigma_{\text{crit}} = \frac{c^2}{4 \pi G} \frac{1}{D_{\mathrm{l}}\beta},
\end{equation}
where $c$ is the speed of light and $G$ is the gravitational constant. The lensing efficiency $\beta$ is defined by
\begin{equation}
\beta = \frac{D_{\text{ls}}}{D_{\text{s}}} H(\zsource-\zlens),
\end{equation}
where $D_{\mathrm{s}}$, $D_{\mathrm{l}}$, $D_{\mathrm{ls}}$ are the angular diameter distances between the observer and the source, the observer and the lens, and the lens and the source, respectively. The Heaviside step function, $H(x)$, is equal to one for positive values of $x$, and zero otherwise. 

In the limit of weak lensing ($\kappa \ll 1$), shape distortions are characterized by the reduced shear $g=g_1+\text{i}g_2$ at position $\btheta$
\begin{equation}
    g(\btheta) = \frac{\gamma(\btheta)}{1-\kappa(\btheta)},
\end{equation}
where $\gamma$ is the (unobservable) complex shear $\gamma = \gamma_1+\rm{i}\gamma_2$ (see, e.g. \citealt{2015RPPh...78h6901K}, for a more detailed account). 

For $|g| \leq 1$, the reduced shear can be estimated from the ensemble-averaged observed ellipticities\footnote{We define ellipticity as $\epsilon = (a-b)/(a+b) \times \mathrm{e}^{\mathrm{2i\phi}}$ for elliptical isophotes with minor-to-major axis ratio $b/a$ and
position angle $\phi$.} $ \epsilon = \epsilon_1 + \mathrm{i} \epsilon_2$, as \citep{1997A&A...318..687S}
\begin{equation}
    \epsilon = \frac{\esource + g}{1 + g^*\esource},
\end{equation}
where $g^*$ denotes the complex conjugate of the reduced shear, and $\esource$ is the intrinsic complex ellipticity of a source galaxy. Because of the intrinsic ellipticities $\esource$, $g$ is not identical to $\epsilon$. However, assuming that the source galaxies have no preferred orientation, the expectation value of $\esource$ vanishes ($ \langle \esource \rangle = 0$), and it holds that $\langle \epsilon \rangle = g$, that is, the ellipticity is an unbiased estimator of the reduced shear.

The dispersion of intrinsic ellipticities is known as shape noise, distinct from uncertainties in measured ellipticities (measurement noise). For simplicity, in this paper we bundle these sources of noise into one entity labelled ``shape noise''. 

The shear, reduced shear and ellipticity can be decomposed into tangential (subscript $\mathrm{t}$) and cross (subscript $\mathsf{x}$) components through 
\begin{subequations}
\label{eq:tancross}
\begin{align}
\label{eq:tancross:tan}
{(\cdot)}_{\mathrm{t}} &= -(\cdot)_1 \cos(2\phi) - (\cdot)_2 \sin(2\phi);\\
\label{eq:tancross:cross}
{(\cdot)}_{\mathsf{x}} &= +(\cdot)_1 \sin(2\phi) - (\cdot)_2\cos(2\phi),
\end{align}
\end{subequations}
where ${(\cdot)}$ denotes any of $g$, $\gamma$ and $\epsilon$, and $\phi$ is the azimuthal angle with respect to a chosen center.
 
For an azimuthally symmetric or azimuthally averaged projected mass distribution, we can write the tangential  shear as a function of projected radius $r$ as 
\citep[e.g.][]{1995ApJ...449..460K,2000ApJ...534...34W}
\begin{equation}
    \gamma_{\mathrm{t}}(r) =  {\bar{\kappa}}(<r) - \kappa(r),
\end{equation}
where ${\bar{\kappa}}(<r)$ is the mean convergence inside $r$.
Equivalently, in terms of the surface mass density,
\begin{equation}
    \gamma_{\mathrm{t}}(r) = \frac{\overline{\Sigma}(<r)-\Sigma(r)}{\Sigma_{\rm{crit}}}.
\end{equation}

\subsection{Simulations}
\label{sec:method:sim}

The analysis in this work is based on the Millennium XXL simulations \citep[henceforth MXXL;][]{2012MNRAS.426.2046A}, starting with cut-outs of massive halos (with a selection described below) from snapshots at $z=0.25$ and $z=1.0$. Particles were extracted from the simulation in a box of 
$3 \times 3 \times 200 ~ (h^{-1} \mathrm{Mpc})^3$  
(comoving) around each halo center (corresponding to the most bound particle in the halo). Shear and convergence images were calculated by projecting particle masses onto a plane. For massive halos, mass distributions were projected along three mutually orthogonal axes to allow for a larger effective sample size. We did not make use of ray-tracing algorithms. Uncorrelated large-scale structure is thus not accounted for in our analysis.\footnote{When comparing to real data, it would be necessary to either include the uncorrelated LSS directly in the simulation, or to add it as a separate component (with zero expectancy) to the error budget. Here care must be taken because the two components may have different radial weights.} 

For comparison, we also made use of the simulations from \citetalias{2011ApJ...740...25B}, in particular the snapshot at $z=0.25$, for comparison with the MXXL snapshot at (approximately) the same redshift. The BK11 data were extracted from a line of sight integration length of 400 $h^{-1}$ Mpc (comoving), twice the value we used for MXXL. In the mass range under consideration, \citetalias{2011ApJ...740...25B} found minimal differences in the mean and the scatter of the weak lensing bias between the two integration lengths considered here. 

The MXXL target halos were originally selected in $\mth$. To also facilitate  an analysis of how the weak lensing bias affects $\mfh$, we completed the sample with a selection down to $\mfh = 1.4 \times 10^{14} h^{-1} M_{\odot}$. In our analysis pertaining to $\mfh$, we subselect halos from the original
$\mth$-selected sample such that the resulting sub-sample is also 95\% complete in $\mfh$.

Because the BK11 sample was selected in $\mfh$, we selected the most massive halos from this simulation, such that the  completeness is greater than $95\%$. The key properties of the simulations, including the number of halos used for estimating the WL bias for $\mfh$ and $\mth$, are summarized in Table~\ref{tab:sims}. 

\begin{table}
\centering
\begin{tabular}{r | l | l | r | r } 
 \hline
 Simulation & snapshot & redshift $z$ & \multicolumn{2}{c}{number of halos} \\
 & & & $M_{200}$ & $M_{500}$ \\
 \hline
MXXL & 41 & 0.989 & 6300 & 4235 \\
MXXL & 54 & 0.242 & 10800 & 7565 \\
BK11 & 141 & 0.245 & 471 & 731 \\
 \hline
\end{tabular}
\caption{Simulations used in this work, and the corresponding number of selected halos for $\mfh$ and $\mth$ in each simulation snapshot.}
\label{tab:sims}
\end{table}

Shear and convergence fields are computed from the projected mass distribution on a grid with a resolution of four arcseconds. The shear fields can be used in different ways to mimic realistic observations. The latter consists of a number of background galaxies, which when ignoring magnification are expected to randomly sample the shear field. We limit the analysis to a radial dependence originating from a constant density of galaxies in the sky plane, which we can adjust to approximately correspond to real observations. We do not take magnification into account, as this is a second-order effect. 

We sample from the shear image to simulate background galaxies at a chosen fixed number density\footnote{By default, galaxies are expected to be randomly positioned on the sky. However, some observational setups may lead to a radius-dependent source density (e.g. \citealt{2018MNRAS.474.2635S}), which can be accounted for in our setup.} (number per angular area on the sky), and transform the shear at each randomly chosen coordinate to reduced shear (ellipticity). In the latter step, for simplicity we choose a constant lensing efficiency $\beta$ corresponding to the mean background source redshift of a given observation.

\subsection{Parametric model}
\label{sec:method:nfw}

Due to the mass-sheet degeneracy \citep{1988ApJ...327..693G, 1995A&A...294..411S}, a direct reconstruction of the projected mass distribution in the sky plane is in general only possible up to a constant. For this reason, one typical approach to determining the mass of a cluster of galaxies with gravitational lensing involves the fitting of a parametric model to a binned profile of tangential reduced shear. A common choice by observers is the NFW profile \citep{1997ApJ...490..493N}, which provides a good match both to DMO \citep{2001MNRAS.321..559B,2012MNRAS.423.3018P,2014ApJ...797...34M,2016MNRAS.457.4340K,2017MNRAS.469.3069G} and hydrodynamical \citep{2014MNRAS.437.2328B,2016MNRAS.456.3542T} simulations. 

While it has been suggested that the Einasto profile \citep{1965TrAlm...5...87E} yields better fits to the mass distributions of simulated halos, in particular at high redshifts (e.g. \citealt{2018ApJ...859...55C}), \cite{2017MNRAS.465.3361H}
showed that this makes little difference for the bias in WL masses. The reason for this is that the central region of clusters is typically excised when fitting to a radial model (e.g. to mitigate the effects of miscentring), effectively eliminating most of the difference between the Einasto and NFW profiles (which are largest close to the center).  

While there are many alternative models for the outer density of a halo (see, e.g., \citealt{2008arXiv0807.3027T,2014ApJ...789....1D}, and references therein), we shall exclusively use the NFW profile in this work as it is the most commonly used model in weak lensing analyses. This does not pose a limitation, as the methods described can be generalized to any radial profile of tangential shear.

The NFW profile is parameterized by 
\begin{equation}
    \rho(r) = \frac{\rho_{\mathrm{crit}}\,\delta_{\mathrm{c}}}{\left(\frac{r}{r_{\mathrm{s}}}\right)\left(1+\frac{r}{r_{\mathrm{s}}}\right)^2},
    \label{nfw1}
\end{equation}
where $r$ is the (three-dimensional) physical radius, $\rho_{\rm{crit}}$
is the critical density of the Universe at the redshift of the cluster, $\delta_{\mathrm{c}}$ is a dimensionless parameter characterizing the density and $r_{\mathrm{s}}$ is a characteristic radius. We define the halo concentration $c_{\Delta}$ as
\begin{equation}
    c_{\Delta} \equiv r_{\Delta}/r_s,
\end{equation}
where $r_{\Delta}$ is the radius within which the mean density is equal to $ \rho_{\rm{crit}} \Delta$, and we employ $\Delta \in \{ \mathrm{200}, \mathrm{500} \} $. The corresponding mass inside $r_{\Delta}$ is given by
\begin{equation}
    M_{\Delta} = \Delta \, \rho_{\rm{crit}} \frac{4 \pi}{3} r_{\Delta}^3.
\end{equation} 
Combining this result with the alternative expression for the mass obtained by integrating \eqn~\eqref{nfw1} leads to
\begin{equation}
\label{eq:rhoc}
    \delta_{\mathrm{c}} = \frac{\Delta}{3} \frac{c_\Delta^3}{f(c_{\Delta})},
\end{equation}
where $f(c_{\Delta}) \equiv \rm{ln}(1+c_\Delta) - c_\Delta/(1+c_\Delta)$. Combining the above equations leads to an expression for the NFW density profile in terms of the mass and the concentration parameter as
\begin{equation}
    \rho(r) = \frac{M_\Delta}{4\pi f(c_\Delta)} \frac{1}{r(r+\frac{r_{\Delta}}{c_{\Delta}})^2}.
\end{equation}
This expression now depends only on mass and concentration, allowing for a relatively simple approach to mass modeling. 

To compare the thus defined NFW model to data, the former must be projected onto the sky plane. The mass surface density is
\begin{equation}
    \Sigma(R) = 2 \int_0^{\infty} \rho \left( \sqrt{R^2+z^2}\right) \text{d} z,
\end{equation}
where $R$ is a projected radius and $z$ is in the direction of the line of sight. 
Exact analytic expressions for the projected surface density and shear of the NFW profile are provided in \cite{1996A&A...313..697B} and \cite{2000ApJ...534...34W}. 

The immediate goal of weak lensing surveys of clusters being the accurate determination of masses, the concentration parameter is often marginalized over in practice. At low signal-to-noise ratios, the degeneracy between the parameters makes it more practical to use a one-parameter fit, in which case a (redshift-dependent) concentration$-$mass relation is used. While many such relations exist, numerical and observational studies alike have found a weak mass dependence of the concentration parameter, as well as a large scatter \citep[e.g.][]{2001MNRAS.321..559B,2008MNRAS.390L..64D,2012MNRAS.423.3018P,2013ApJ...766...32B,2014MNRAS.441.3359D,2014MNRAS.441..378L,2015ApJ...799..108D,2016MNRAS.460.1214L,2017ApJ...840..104S,2019ApJ...871..168D,2019MNRAS.486.4001R}.

As long as the mass bias can be determined accurately from simulations, it does not matter in principle what concentration$-$mass relation is used; choosing a suitable constant value for $c_{\Delta}$ is also a valid approach as long as data and simulations are treated equivalently. In practice, a useful consideration amounts to finding an operating point at which the combined systematic and statistical uncertainty of the determined mass is as small as possible. As we shall see in \sect~\ref{sec:res:biasdep}, this choice is critically dependent on the radial range of the fit as well as on the mass and redshift, in addition to being dependent on the specific concentration$-$mass relation under consideration.

\subsection{Adding noise}
\label{sec:meth:addnoise}

As one aim of this paper we want to investigate whether conditions exist under which the weak lensing mass bias is independent of the absolute noise level. We describe here how we add noise to the simulated shear measurements. Because the shear images are extracted from the simulations, correlated LSS is already included. We thus add random shape noise, with zero mean and with variance $\sigma_{\mathrm{s}}^2$, to each tangential shear bin according to 
\begin{equation}
    \label{eq:addnoise}
    \sigma_{\mathrm{s}}^2 = \frac{\sigma_{\mathrm{e}}^2}{n_{\rm{gal}} A},
\end{equation}
where $\sigma_{\mathrm{e}}$ is the intrinsic shape noise, which we assume to be the same for all lensed sources, $n_{\rm{gal}}$ is the surface density of background galaxies on the sky, and $A$ is the angular area of the annulus. For simplicity we assume a constant surface density, that is, we disregard magnification as well as observational effects such as blending by cluster galaxies. In general, we assume an intrinsic shape noise of $\sigma_{\mathrm{e}} = 0.25$, which is close to typical values for both ground-based and space-based observations.  

In our method, the tangential ellipticies of the galaxies are given equal weights, and only the number of galaxies in a bin is important for the uncertainties of the binned data. While it is also possible to fit directly without binning the ellipticities, \cite{2012MNRAS.421.1073B} showed explicitly that this choice of method has no bearing on the determination of the WL mass bias. 

\subsection{Mass bias modeling}
\label{sec:meth:massbias}

For forward modeling of a given weak lensing observation, we seek to determine the probability of measuring a weak lensing mass $\wlmass$ by fitting a shear profile to a radial model given a true mass $\truemass$. That is, we seek to determine the probability $P(\wlmass|\truemass)$.
We define the linear bias as
\begin{equation}
\label{eq:bias_definition}
\bias \equiv \frac{\wlmass}{\truemass},
\end{equation}
where both masses are evaluated at the same spherical ovendensity $\Delta$. It is important to realize that while for an individual halo $\bias$ is just a number, our goal is to determine the distribution of $\bias$ from an ensemble of simulated halos. Traditionally, a log-normal distribution has been assumed (e.g. \citealt{2019MNRAS.483.2871D}, \citealt{2018MNRAS.474.2635S}, \citealt{2018MNRAS.479..890L}), i.e.
\begin{equation}
    \label{eq:lognormal0}
    \ln \left(\frac{\wlmass}{\truemass}\right) \sim \mathcal{N}(\ln \mu,\sigma^2),
\end{equation}
where $\mathcal{N}(\ln \mu,\sigma^2)$ is the normal distribution with mean $\ln \mu$ and variance $\sigma^2$. Note that with this definition, the variance $\sigma^2$ is native to log-space, while the mean $\mu$ is native to the linear space of $\bias$. In this sense, $\mu$ is related to the expectation value of $\bias$, but the full distribution must be taken into account in modeling the bias of actual observations; $\mu$ is not generally the mean of $\bias$ for a sample of halos.

Given a set of simulations with known $\truemass$ and making the log-normal assumption, the task of finding $P(\wlmass|\truemass)$ becomes estimating the probability
\begin{equation}
\label{eq:main_prob}
    P(\mu,\sigma|\hat{g}) \propto P(\hat{g} | \mu,\sigma) P(\mu,\sigma)
\end{equation}
where $\hat{g}$ is the observed reduced shear, and $P(\mu,\sigma)$ is the prior on the parameters of the log-normal distribution. We use a top-hat prior with $0.01 < \sigma < 10$ and $0.5 < \mu < 2$. 

\subsubsection{Single halo}

We consider first a single simulated halo. Marginalizing the factor $P(\hat{g} | \mu,\sigma)$ over $\wlmass$, we may write
\begin{equation}
\begin{split}
    \label{eq:p_of_g_given_lognormal}
    P(\hat{g} | \mu,\sigma) &\propto \int_{\wlmass} P(\hat{g},\wlmass|(\mu,\sigma)) ~\text{d} \wlmass \\
     &\propto \int_{\wlmass} P(\hat{g}|\wlmass) P(\wlmass|(\mu,\sigma)) ~\text{d} \wlmass,
\end{split}
\end{equation}
where we have also used the fact that given $\wlmass$, $\hat{g}$ and $(\mu,\sigma)$ are conditionally independent so that $P(\hat{g}|\wlmass,(\mu,\sigma)) = P(\hat{g}|\wlmass)$.

The first probability inside the integral of \eqn~\eqref{eq:p_of_g_given_lognormal} is obtained 
from fitting the simulated reduced shear profile with the model prediction given $M_\mathrm{WL}$. Specifically, 
\begin{equation}
    P(\wlmass|\hat{g}) \propto P(\hat{g}|\wlmass)P(\wlmass),
\end{equation}
where the prior $P(\wlmass)$ does not need to be the same for all halos in the sample, but needs to be chosen carefully as we shall see below. We can use different approaches for estimating $P(\hat{g}|\wlmass)$: sampling by Markov-Chain Monte Carlo (MCMC) or using a grid-search on $\wlmass$ with step $\delta \wlmass$ in linear mass (if a concentration$-$mass relation is assumed). While both approaches are equally valid, we use MCMC samplig in this work.

The second probability in the integrand is the log-normal distribution, which is given by \eqref{eq:lognormal0}. Explicitly, this probability is
\begin{equation}
    P(\wlmass| (\mu,\sigma)) = \frac{1}{\bias \sqrt{2 \pi \sigma^2}} \exp{\left(-\frac{ \left( \ln \bias - \ln \mu \right)^2}{2 \sigma^2}\right)}.
\end{equation}
To evaluate the integral over the product of the two probabilities, we use the approximation
\begin{equation}
    \label{eq:pdf_integral}
    P(\hat{g} | \mu,\sigma)_{\rm{grid}} \appropto \sum_{k=1}^{N_p} P(M_k| \mu,\sigma) P(M_k|\hat{g}) \frac{1}{P(M_k)} \delta \wlmass
\end{equation}
for the grid-search method with $N_p$ the number of sampled points and $\delta \wlmass$ the mass step size of the grid, and
\begin{equation}
    \label{eq:mcmc_integral}
    P(\hat{g} | \mu,\sigma)_{\rm{MCMC}} \appropto \frac{1}{N_s} \sum_{k=1}^{N_s} P(M_k| \mu,\sigma) \frac{1}{P(M_k)} M_k
\end{equation}
for the MCMC method with $N_s$ the number of samples in the Markov Chain. Here, $M_k$ is the weak lensing mass corresponding to sample $k$.

We call attention to a few subtleties at this point. First, the sampling in weak lensing mass does not need to extend to negative masses\footnote{Negative masses are permissible in principle, namely as noisy measurements of mass when the relative mass uncertainty is comparable to one.}, as the log-normal part of the integrand is undefined for such masses. We explore the validity of the log-normal assumption in \sect~\ref{sec:results}. Second, because the prior $P(\wlmass)$ is taken into account in \eqns~\eqref{eq:pdf_integral} and \eqref{eq:mcmc_integral}, we can choose it freely when fitting for $P(\hat{g}|\wlmass)$. We discuss this point in some detail in \sect~\ref{sec:meth:massbias:prior}. 

\subsubsection{Sample of halos}

Having worked out the formalism for a single halo, we move on to a sample of $N_c$ halos. We define the likelihood $\mathcal{L}(\mu,\sigma)$ for an ensemble of halos from \eqn~(\ref{eq:main_prob}) 
\begin{equation}    
P(\mu,\sigma|\hat{g}) \propto P(\hat{g}|\mu,\sigma) P(\mu,\sigma) \equiv \mathcal{L}(\mu,\sigma) P(\mu,\sigma),
\end{equation}
where
\begin{equation}
    \mathcal{L}(\mu,\sigma) = \prod_{i=1}^{N_c}  P_i(\hat{g_i}|(\mu,\sigma)),
\end{equation}
where $P_i(\hat{g_i}|(\mu,\sigma))$ is given for an individual halo by \eqns~\eqref{eq:pdf_integral} and \eqref{eq:mcmc_integral}. Now let $M_{ij}$ be the $j$th MCMC sample of weak lensing mass from halo $i$, and $b_{ij}$ the corresponding linear bias. For the MCMC method we shall then have
\begin{align}
\label{eq:finalfit}
    \nonumber \mathcal{L}(\mu,\sigma) =& \prod_{i=1}^{N_c} \Bigg\{ \frac{1}{N_s(i)} \sum_{j=1}^{N_s(i)} \Bigg[\frac{M_{ij}}{P(M_{ij})} \frac{1}{b_{ij}\sqrt{2 \pi \sigma^2}} \times \\
    & \exp \left( {-\frac{(\ln b_{ij}- \ln \mu)^2}{2 \sigma^2}} \right) \Bigg] \Bigg\},
\end{align}
where $N_s(i)$ is the number of samples for halo $i$ (which does not necessarily need to be the same for all halos). 

\subsubsection{Choice of mass prior}
\label{sec:meth:massbias:prior}

Having derived the basic formalism for the likelihood of a given log-normal distribution of the bias, we turn our attention to the choice of the mass prior, $P(\wlmass)$. At high statistical noise, some of the simulated halos will up-scatter enough that the overlap between $P(\hat{g}|\wlmass)$ and $P(\wlmass|(\mu,\sigma))$ becomes very small. Because the extreme tails of a distribution are never sampled by a Markov chain with a finite number of steps, this can lead to the likelihood contribution from these targets being severely underestimated, with the end result of overestimating $\sigma$ and underestimating $\mu$. An uninformed prior on mass is therefore not ideal, and would require millions of samples per halo even at signal-to-noise ratio levels of $2-3$ (while many halos in our analysis in fact have signal-to-noise ratios less than 2).  

While a prior proportional to inverse mass (corresponding to an uninformed prior on the logarithm of mass) may somewhat relieve this problem, we have found that this is not optimal. Instead we use the information on true masses and define the prior in terms of the bias $\bias$. Because we multiply the resulting distribution with a log-normal, we have found that a prior that is itself log-normal works quite well. While $\mu$ and $\sigma$ are not known a priori, we can still identify a relevant range for these parameters and choose the log-normal prior accordingly. In \sect~\ref{sec:res:prior_robust} we describe a method to test for the robustness of the mass prior. We set 
\begin{equation}
    P(\ln \bias) \sim \mathcal{N}(\ln \mu_{\rm{prior}},\sigma_{\rm{prior}}^2),
\end{equation}
which for a halo with index $i$ can be converted to a prior on weak lensing mass $M_i$ using \eqn~\eqref{eq:bias_definition} and the knowledge of the true mass ${\truemass}_{i}$. We have found that the choice $(\ln \mu_{\rm{prior}}, \sigma_{\rm{prior}}) = (0.0,0.5)$  works well for typical bias distributions reported in the literature (e.g. \citealt{2011ApJ...740...25B,2018MNRAS.479..890L}). With 10000 samples per halo, we can accurately reproduce input distributions in the range $-0.2 < \ln \mu  < 0.2$ and $0.1 < \sigma <  0.4 $ as discussed in \sect~\ref{sec:res:prior_robust}.

\subsection{Miscentring distributions}
\label{sec:meth:miscentring}

We explain here how we derive empirical noise-dependent 
miscentring distributions from the peaks of recovered convergence signal-to-noise ratio (henceforth SNR) images. 
This analysis is restricted to the $z=1$ snapshot of the MXXL simulation\footnote{We do not consider the redshift dependence of the miscentring distribution in this work.}, using a lensing efficiency of 0.3. We use an idealized square field of view with a side of 6.4 arcminutes, approximately mimicking a 2$\times$2 mosaic with HST-ACS, and a shape noise $\sigma_e=$0.25 for each lensed galaxy.

We construct shear catalogs with different noise levels by varying the galaxy number density. For the main analysis, we use a value of 20 arcmin$^2$ for all fields, which approximately matches the setup for two-filter HST/ACS mosaics and clusters in the redshift interval $0.7<z<1$ in \cite{2020arXiv200907591S}.

For the convergence reconstruction, we use a grid-based Wiener filter approach as described by \citet{2009MNRAS.399L..84M} and \citet{2009MNRAS.399...48S}. We use the implementation from the latter reference, employing the measured ellipticity two-point correlation function (e.g. \citealt{2001PhR...340..291B}) for the computation of the Wiener filter for each halo. 

While the convergence can only be determined up to a constant due to the mass-sheet degeneracy (see \sect~\ref{sec:method:nfw}), this has no bearing upon the miscentring distribution as it does not change the position of peak in the SNR image. Because it does play a role in the determination of the SNR at the peak, however, we set the average convergence in each target field to zero. 

We randomize the data in three different ways in order to derive SNR images:

\begin{enumerate}
    \item Randomization by phase. This approach rotates the phase of each ellipticity by adding a random angle between 0 and $\pi$.
    \item Randomization by position. Because we do not take magnification into account and have a constant galaxy density across the field, we can also bootstrap by randomly changing the position of each galaxy in the shear catalog. 
    \item Randomization by galaxy selection. This approach differs from the previous two in that it preserves the halo signal. It works by randomly sampling each shear catalog with replacement (bootstrapping), as often employed in lensing analyses (e.g. \citealt{2018MNRAS.474.2635S}). 
\end{enumerate}

Note that (i) and (ii) yield noise images, through which the actual reconstructions are divided. For each of the three randomization schemes, we make 400 randomized images of each field. In addition, we make 400 independent noise realizations of each field, to serve as a reference. In total, we thus make 1600 convergence reconstructions for each target. In order to reduce the computing time, we only process every fourth halo (in the order of descending $M_{200}$). 

SNR images were made, for each randomization method, by computing the standard deviation in each $4^{\prime\prime} \times 4^{\prime \prime}$ image pixel across all 400 realizations (also for method (iii)). We find that the randomization method plays a very small role in determining the SNR, as reported in \sect~\ref{sec:res:kappamis:snr}. From here on, we work with positionally randomized ellipticities. 

We searched for the peak SNR inside a two arcminute radius from the known simulation center. This approach is not quite realistic, as there is no such known starting point in real observations. We discuss this problem in \sect~\ref{sec:disc:kappa}. Miscentring distributions, with respect to true halo centers, were derived from the recovered convergence SNR peaks, binning by SNR at the peak position. 

At this point, we also investigated whether the SNR images produced by bootstrapping can be used to get a reliable estimate of the uncertainty in the convergence peak position. 
In each field, we measured the mean and median offset from the nominal position for each of the 400 bootstrapped realizations, and compared these values to those obtained from the reference (independent noise realizations) over the full sample as a function of SNR. 

\section{Results}
\label{sec:results}

We begin this section by showing that the chosen mass prior is robust. Using that information, we show that with our simulation setup, the resulting bias distribution is noise independent, allowing us to directly estimate the bias distribution and test the validity of the assumption of this distribution being log-normal. 
We then move on to investigating the dependence on radial range, on the concentration$-$mass relation and on miscentring. We show that when including miscentring in the bias estimation, the resulting distribution is far from log-normal. 

\subsection{Robustness of the mass prior}
\label{sec:res:prior_robust}

The mass prior is based on the true mass of each halo in the simulation, as described in \sect~\ref{sec:meth:massbias}. To test the robustness of this mass prior and to contrast it with an inverse mass prior, we carry out a set of simplistic Monte Carlo simulations. 

We draw artificial samples from a hypothetical sample of 50,000 halos at fixed mass $M_{0}$. To mimic a mass bias, each individual mass is first offset from the nominal mass using a log-normal distribution (\sect~\ref{sec:meth:massbias}) with  
$(\ln \mu_0, \sigma_0) = (0.0,0.5)$. 
Additionally, we impose a measurement error on the biased mass using a normal distribution with mean zero and variance $\sigma_f^2 = f \, M_0$, where $f$ is varied in the range 0.01 to 2.  
For each halo, we draw 1000 samples centered on the ``measured'' mass (the ``best-fit'') from a Normal distribution with variance $\sigma_f$. With these data, we repeatedly use \eqn~(\ref{eq:finalfit}) with different mass priors and different values of $f$. 

To quantify the robustness of the priors in terms of how well we can reproduce $\sigma_0$ and $\mu_0$ using the likelihood function given by \eqn~(\ref{eq:finalfit}), we compute the relative quantities
\begin{equation}
    \label{eq:murel}
    \theta_{\mu} = \frac{\mu - \mu_0}{\mu_0} 
\end{equation}
and
\begin{equation}
    \label{eq:sigmarel}
    \theta_{\sigma} = \frac{\sigma - \sigma_0}{\sigma_0}.
\end{equation}

We test three different priors: a top-hat prior, an inverse mass prior (corresponding to an uninformed prior in the logarithm of the mass), and the log-normal prior, based on the known masses as described in \sect~\ref{sec:meth:massbias:prior}. We judge the  merit of each prior based on a 5\% systematic deviation in the mean $\mu$ of the bias distribution $\bias$. 

With increasing fractional uncertainty $f$, we find that $\mu$ increases and eventually diverges, as $\sigma$ decreases and similarly diverges. The top-hat mass prior performs poorly, with a systematic error in $\mu$ of $-$5\% at $f=0.4$, and diverging at higher values of $f$. Correspondingly, the scatter $\sigma$ is overestimated by a relative +8\% at $f=0.4$. The inverse mass prior slightly improves the situation, with a corresponding systematic of ($-$5\%,+8\%) in ($\mu$,$\sigma$) occurring approximately at $f=0.8$. The most robust results are achieved using the log-normal mass prior, where the systematic errors in $\mu$ and $\sigma$ remain below 1\% and 2\%, respectively, at $f=1$ (corresponding to a signal-to-noise ratio of 1)\footnote{This holds under the assumption that the underling distribution is indeed log-normal. Distributions with wide tails would require another approach, such as importance sampling.}. We use this mass prior in the following.

\subsection{Noise level independence}
\label{sec:res:noise_indep}

We move on to show that the mass bias distribution is independent of the absolute noise level of the simulated measurements, under the assumption of an underlying log-normal distribution. To this end, we divide the MXXL sample at $z=1$ into mass bins. Because the samples are not uniformly distributed in mass, we define the mass bins so as to include similar numbers of halos. The mass bins are listed in Table \ref{tab:massbins}. 

\begin{table}
\centering
\begin{tabular}{r | l | l } 
Bin no. & $\log_{10}(\mth [M_{\odot}])$ & $\log_{10}(\mfh [M_{\odot}])$  \\
\hline
\multicolumn{3}{c}{{MXXL at $z=1.0$}} \\
\hline
0 & 14.68$-$14.71 (1323) & 14.28$-$14.37 (531) \\
1 & 14.71$-$14.73 (840) & 14.37$-$14.60 (512) \\
2 & 14.73$-$14.76 (1119) & 14.60$-$14.64 (975) \\
3 & 14.76$-$14.80 (933) & 14.64$-$14.69 (864) \\
4 & 14.80$-$14.85 (819) & 14.69$-$14.85 (1110) \\
5 & 14.85$-$15.00 (1062) & 14.85$-$15.20 (243) \\
6 & 15.00$-$15.28 (204) &  $-$ \\
\hline
\multicolumn{3}{c}{{MXXL at $z=0.25$}} \\
\hline
0 & 14.31$-$14.55 (600)  & 14.28$-$14.43 (589) \\
1 & 14.55$-$14.78 (600) & 14.43$-$14.44 (590) \\
2 & 14.78$-$14.95 (600) & 14.44$-$14.55 (225) \\
3 & 14.95$-$15.02 (1428) & 14.55$-$14.82 (210) \\
4 & 15.02$-$15.05 (1602) & 14.82$-$14.87 (612) \\
5 & 15.05$-$15.09 (1734) & 14.87$-$14.92 (1076) \\
6 & 15.09$-$15.14 (1521) & 14.92$-$14.96 (1293) \\
7 & 15.14$-$15.20 (1119) & 14.96$-$15.03 (1416) \\
8 & 15.20$-$15.30 (981) & 15.03$-$15.13 (930) \\
9 & 15.30$-$15.40 (411) & 15.13$-$15.20 (348) \\
10 & 15.40$-$15.74 (204) & 15.20$-$15.56 (276) \\
\hline
\multicolumn{3}{c}{{BK11 at $z=0.25$}} \\
\hline
0 & 14.70$-$14.80 (233) & 14.45$-$14.55 (322) \\
1 & 14.80$-$14.92 (141) & 14.55$-$14.75 (308) \\
2 & 14.92$-$15.34 (97) & 14.75$-$15.10 (101) \\
\end{tabular}
\caption{Mass bins used in the analysis. Numbers in parentheses indicate the number of halos in each bin. The bins in $\mth$ and $\mfh$ are defined independently, and bins with the same bin number do not necessarily correspond to recomputing the mass limits from one over-density to the other.}
\label{tab:massbins}
\end{table}

We estimate the bias distribution for each mass bin at different noise levels, as described in \sect~\ref{sec:meth:addnoise}. In the particular case of a noiseless realization, we cannot use \eqn~\eqref{eq:addnoise}. Instead, we use weights to ensure that the radial bins of reduced shear are weighted the same way as in the presence of noise. We quantify the noise level relative to a reference level, defined as having a surface density $N_\text{\rm{gal}} = 10~\rm{arcmin}^{-2}$ with a shape noise $\sigma_{\mathrm{e}}=0.25$ at a lensing efficiency $\beta=0.3$.    

At zero noise, we can test for log-normality in the bias distribution of each mass bin by directly applying \eqn~\eqref{eq:bias_definition} with no need for fitting for the distribution parameters. 

\begin{figure}	
\includegraphics[width=\columnwidth,clip=True,trim={0 0 0 0}]{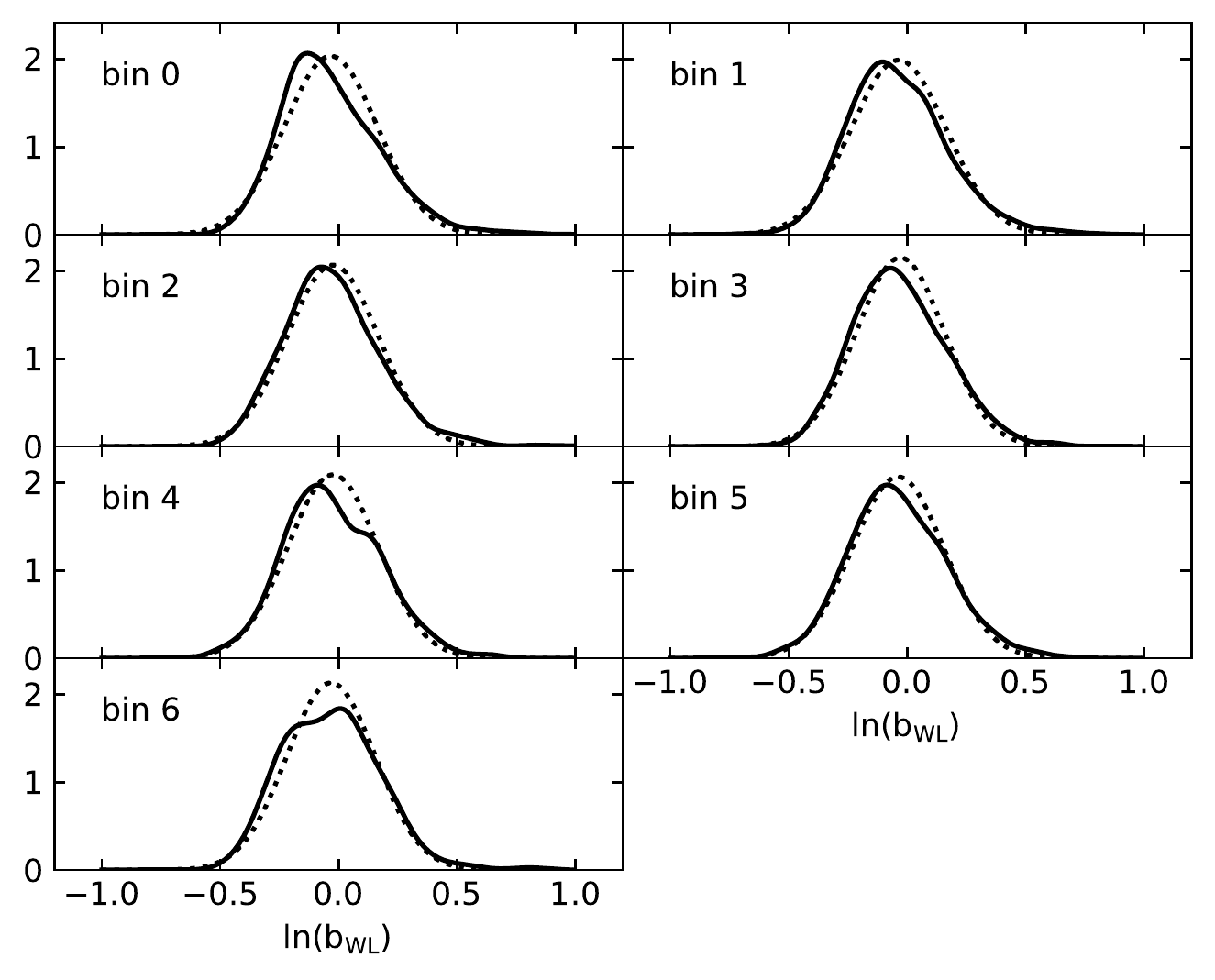}    \\
\includegraphics[width=\columnwidth,clip=True,trim={0 0 0 0}]{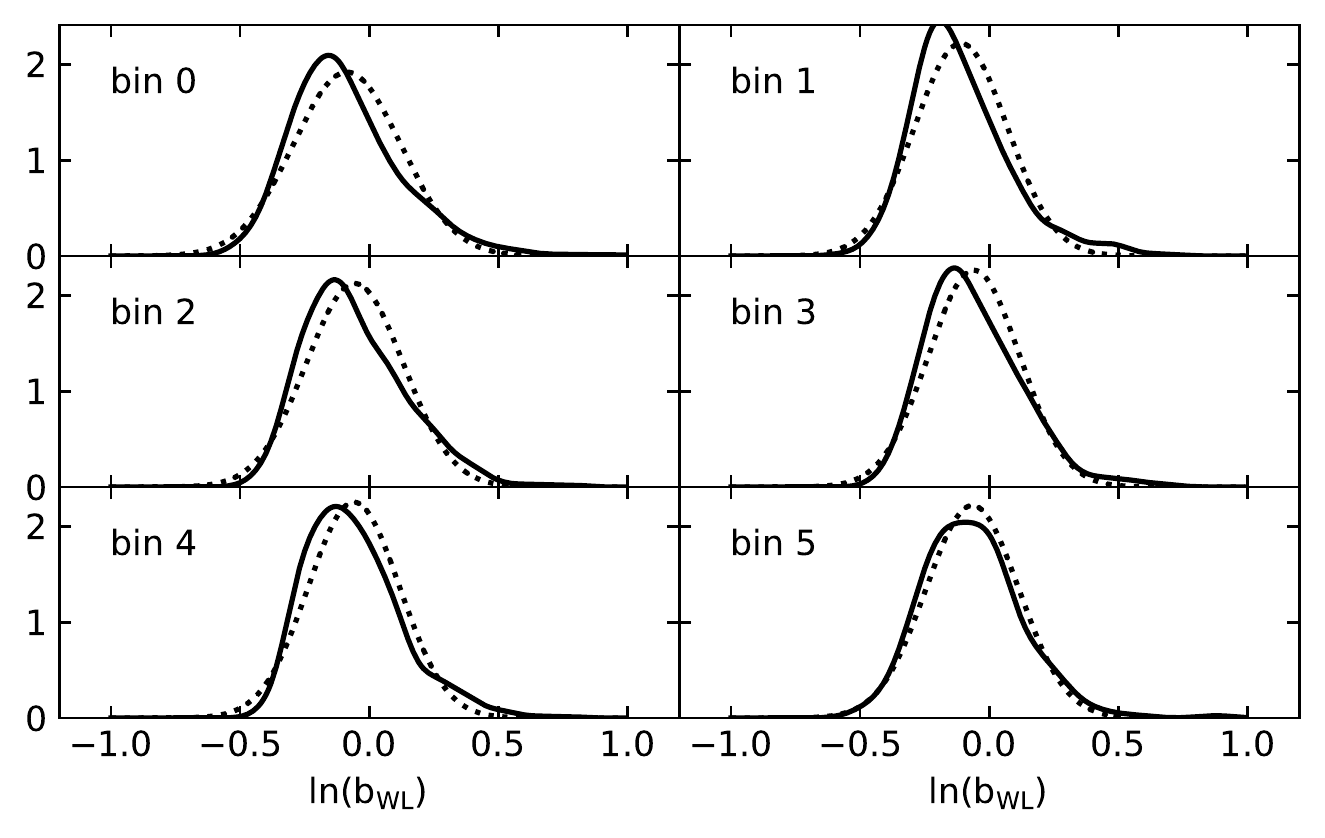}  
\caption{Actual versus re-sampled bias distributions at zero noise in bins of $\mth$ (top) and $\mfh$ (bottom) for the $z=1$ MXXL snapshot. Solid lines indicate the actual distributions; dotted lines represent the corresponding log-normal distributions with the same mean and variance.}
\label{fig:isitlognormal}
\end{figure}

As the distributions do not appear exactly log-normal (Fig.~\ref{fig:isitlognormal}), we re-sample the halos in each mass bin to mimic log-normal distributions. In particular, we construct an empirical, parameter-free model of each distribution and sample from it so as to obtain the largest possible sample of halos consistent with a log-normal distribution. The latter is defined by the sample mean and sample variance of the measured distribution at zero noise. 

\begin{figure*}	
\includegraphics[width=0.48\textwidth,clip=True,trim={0 0 0 0}]{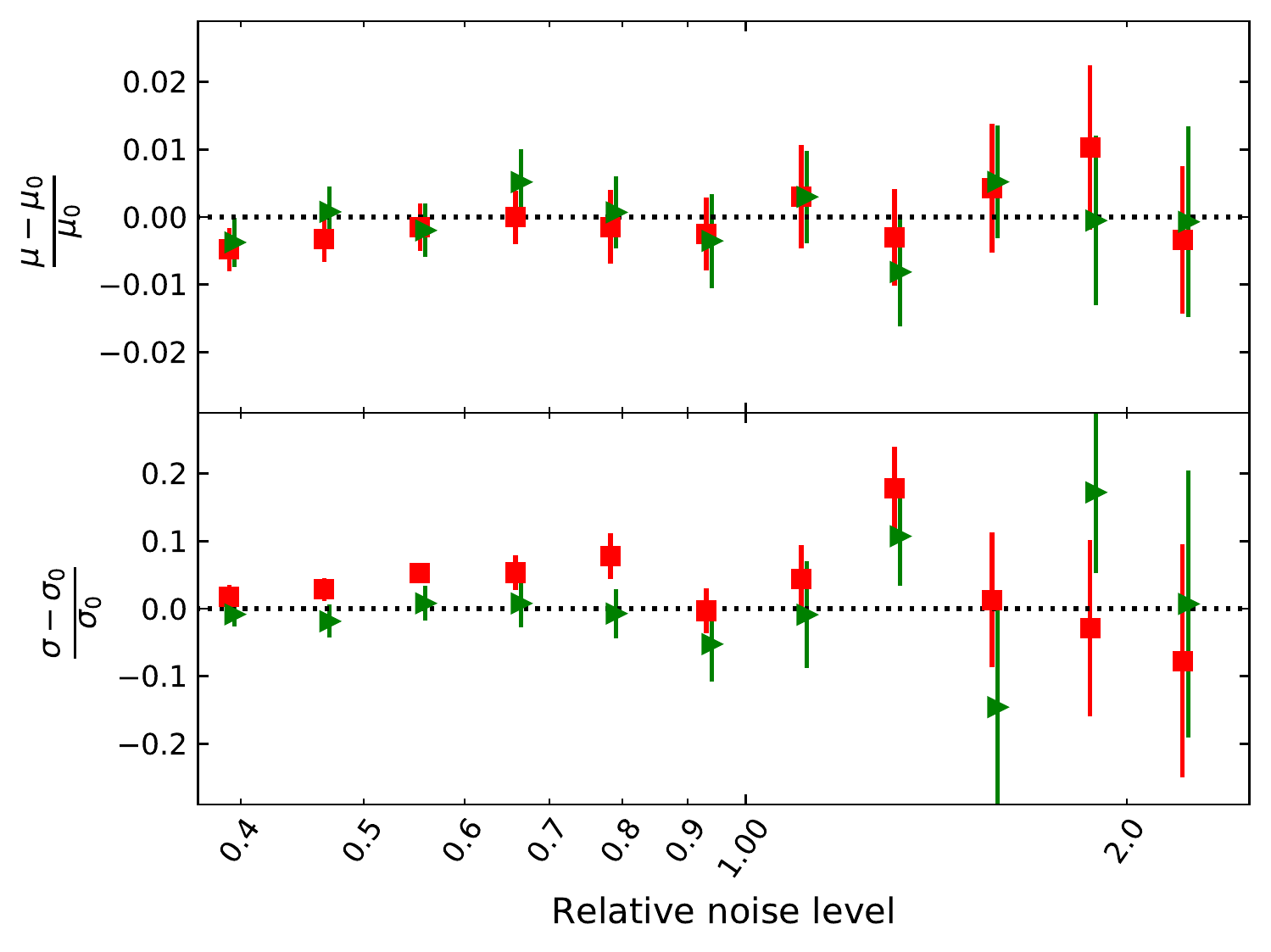}  
\includegraphics[width=0.48\textwidth,clip=True,trim={0 0 0 0}]{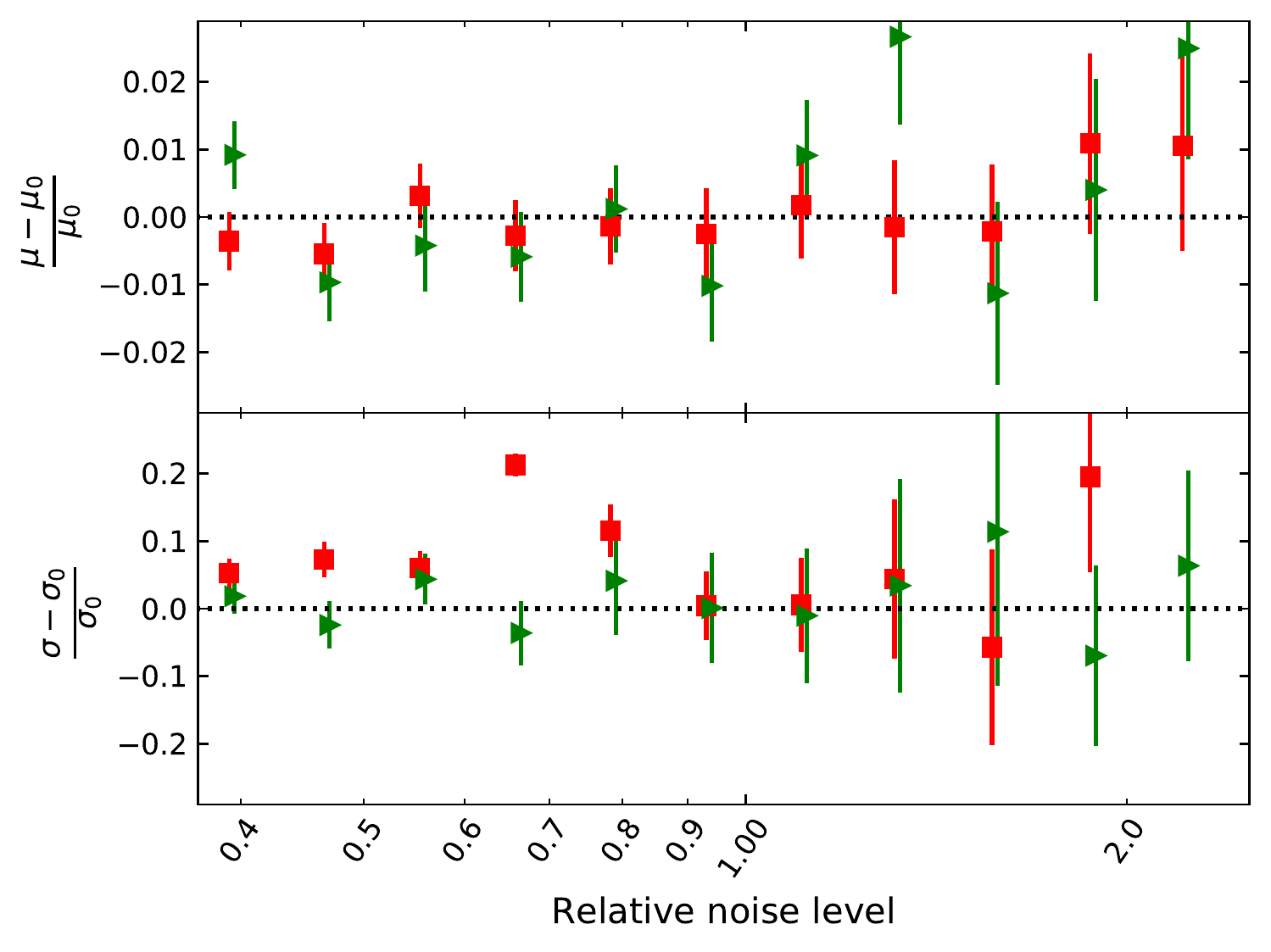}  
\caption{Relative bias in the two parameters of the estimated bias distribution, using the MXXL snapshot at $z=1$. Results were combined for all mass bins at each relative noise level. A noise level of 1 means a background galaxy surface density of 10 arcmin$^{-2}$ and a shape noise of 0.25. Green data points were generated with an underlying log-normal distribution (see text). Red data points were generated with the original underlying bias distribution of the simulation. Left: $\mth$. Right: $\mfh$.}
\label{fig:lognormal.no.bias}
\end{figure*}

At each noise level and for each mass bin, we construct the relative quantities $\theta_\mu$ and  $\theta_{\sigma}$, defined in \sect~\ref{sec:res:prior_robust}, where the reference values $\mu_0$ and $\sigma_0$ now come from the log-normal distribution constructed from the noiseless case as described above. At a given noise level, we combine all $\theta$ by considering their uncertainties as two-sided Gaussians. The results are shown in Fig.~\ref{fig:lognormal.no.bias}. Our results are consistent with no additional bias in the distribution parameters $\mu$ and $\sigma$, provided that the underlying distribution is log-normal. Using the original underlying distributions, which are not perfectly log-normal, results in an over-estimation of $\sigma$ of up to one tenth of its value at low noise levels for both $\mth$ and $\mfh$. Given typical levels of $\sigma$, the result is not significant. The mean bias $\mu$ is consistent to within 1\% when the original distributions are used. 

Our results suggest that the bias determination is independent of the absolute noise level, provided that the bias distribution is log-normal. This has the fortunate side effect that we can model the distribution directly using noiseless simulations. We will make use of this in the next subsection. In order to rule out that this result is a statistical fluke, we have verified it for all the results presented in the following. 

\subsection{Mass bias dependencies}
\label{sec:res:biasdep}

While the WL bias distribution is independent of the absolute noise level, it is dependent on a number of factors, such as mass and redshift, the radial range in which the NFW profile is fit, and the choice of concentration$-$mass relation. We investigate some of these dependencies here in order to show some general trends, and to give a general idea of how the data analysis can be optimised so as to minimize the bias amplitude (deviation from 1) and scatter in the WL bias. 

To this end, we start by defining a fiducial setup, from which we then deviate in a number of ways to investigate the general dependencies. Our fiducial setup uses the concentration$-$mass relation of \citet{2015ApJ...799..108D}, with the corrected parameter set of \citet{2019ApJ...871..168D}. The radial range of the fit is from $r_{\rm{min}}=0.5$ Mpc to $r_{\rm{max}}=3.5$ Mpc  (physical), with 15 radial bins. Based on the results of \sect~\ref{sec:res:noise_indep}, we use noiseless simulations. We set the lensing efficiency $\beta$ to a constant value of $0.3$ at $z=1$ and $0.7$ at $z=0.25$. 

\subsubsection{Mass and redshift}

\cite{2011ApJ...740...25B} modeled the mass bias from a set of simulated weak lensing observations as a power law, with the independent variable being the true mass and the dependent variable the measured mass, and included a log-normal scatter term. We follow a similar approach here. However, while this method naturally allows for a first-order estimate of a mass dependence in the bias (through a slope different from unity), we seek to have a more flexible constraint on the mass dependence, and model it in discrete bins instead. In that sense, our method is similar to that of \cite{2018MNRAS.479..890L}, although we do not allow the concentration parameter to vary freely. 

\begin{figure*}	
\includegraphics[width=0.48\textwidth,clip=True,trim={0 0 -10 5}]{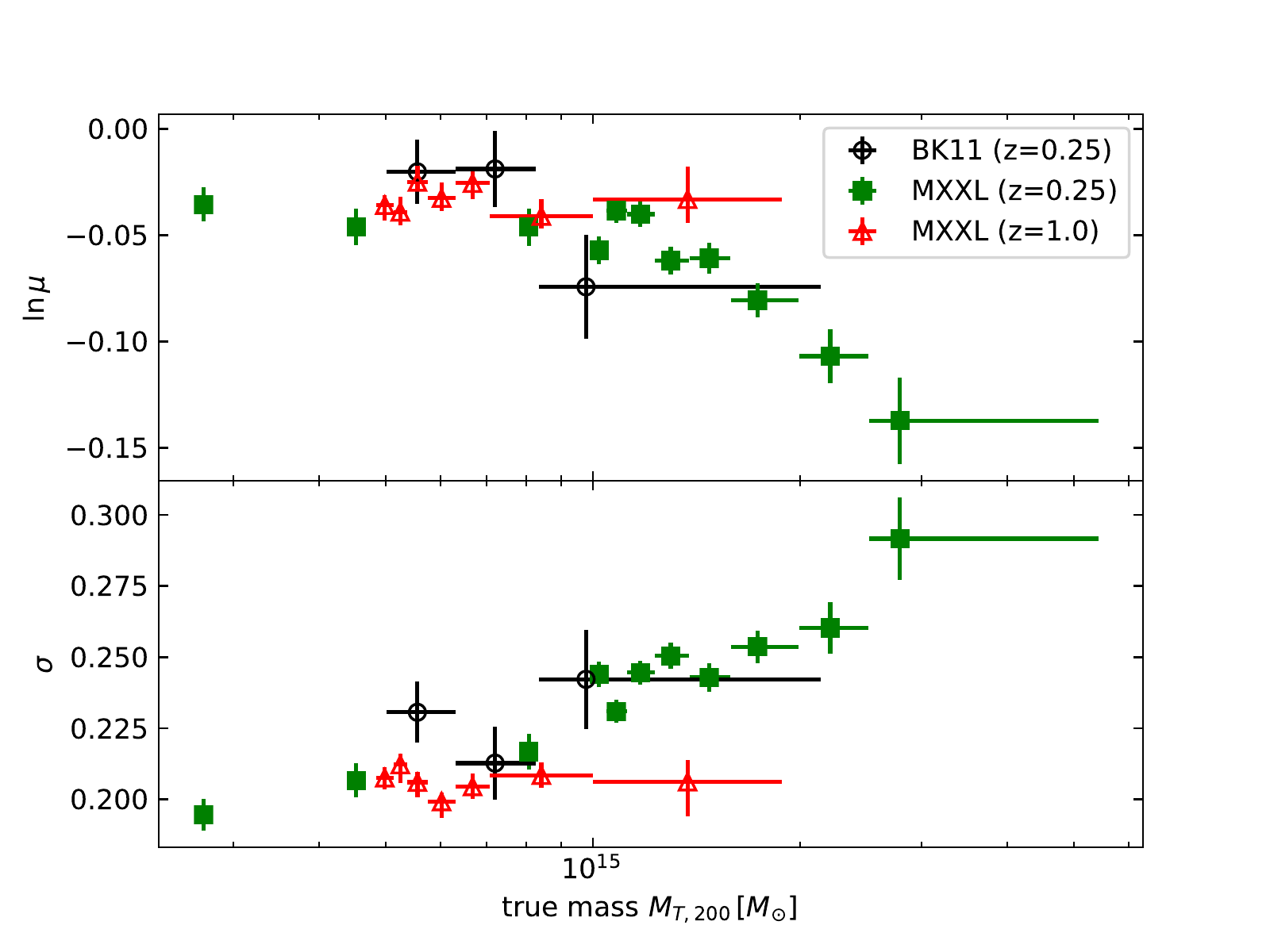} 
\includegraphics[width=0.48\textwidth,clip=True,trim={0 0 -10 5}]{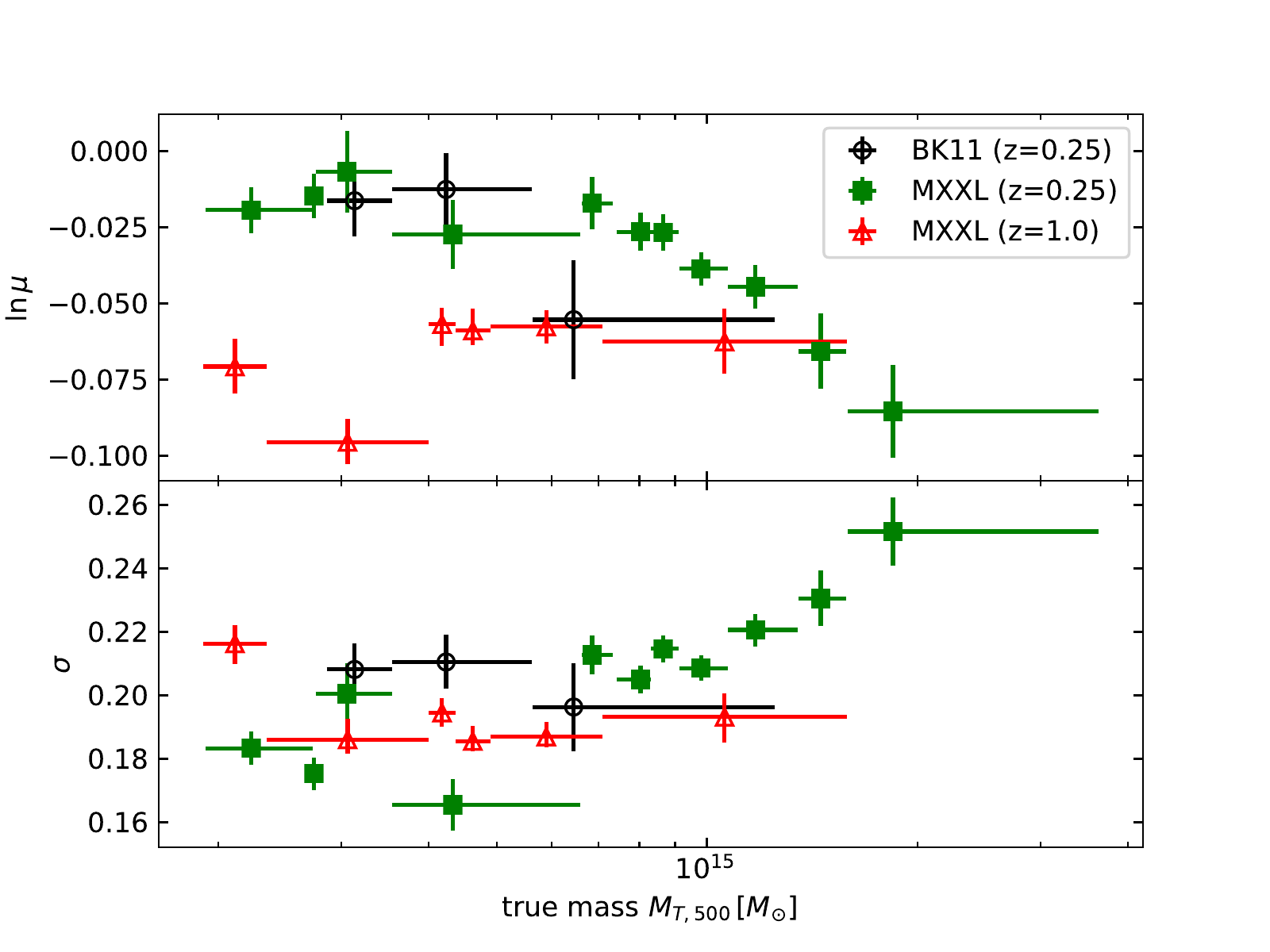}
\caption{Log-normal bias distributions from the BK11 and MXXL simulations at $z=0.25$, and for the MXXL simulation at $z=1.0$. Simulated halos were binned by the true mass. For these results, the default setup was used, with the concentration$-$mass relation of \citet{2015ApJ...799..108D} (corrected parameter set of \citet{2019ApJ...871..168D}), the radial range $0.5-3.5$ Mpc  (physical), and no miscentring. Left: $\mth$. Right: $\mfh$.}
\label{fig:biasdep:bk11mass}
\end{figure*}

Figure \ref{fig:biasdep:bk11mass} shows the mass dependence
of the bias parameters $\mu$ and $\sigma$ in three different simulation snapshots. First, we compare MXXL snapshot 54 to BK11 snapshot 141 (both at $z=0.25$). The MXXL snapshot was divided into 11 bins in {$M_{200}$} and {$M_{500}$}, while the smaller BK11 snapshot was divided into 3 mass bins for both over-densities. The direct comparison is limited by the number of targets in the BK11 simulation, resulting in a fractional uncertainty in $\mu$ of $2.4\%$ for the bin with the highest $\mth$. Within this uncertainty, the results are in reasonable agreement. For $\mfh$, the scatter $\sigma$ is somewhat lower in the MXXL simulation.

We also show a comparison of the two redshift slices of the MXXL simulations in Fig.~\ref{fig:biasdep:bk11mass}. Investigating the $z=0.25$ snapshot for our fiducial analysis, we find that mass estimates are biased low more strongly at higher masses compared to lower masses, while the scatter of the bias distribution increases with mass. At redshift $1$, this trend vanishes. However, as we shall see in \sect~\ref{sec:res:biasdep:mc}, this is more a consequence of the chosen concentration$-$mass relation than a statement about WL bias in general.

\subsubsection{Radial range}

\begin{figure*}	
\includegraphics[width=0.48\textwidth,clip=True,trim={0 0 -10 5}]{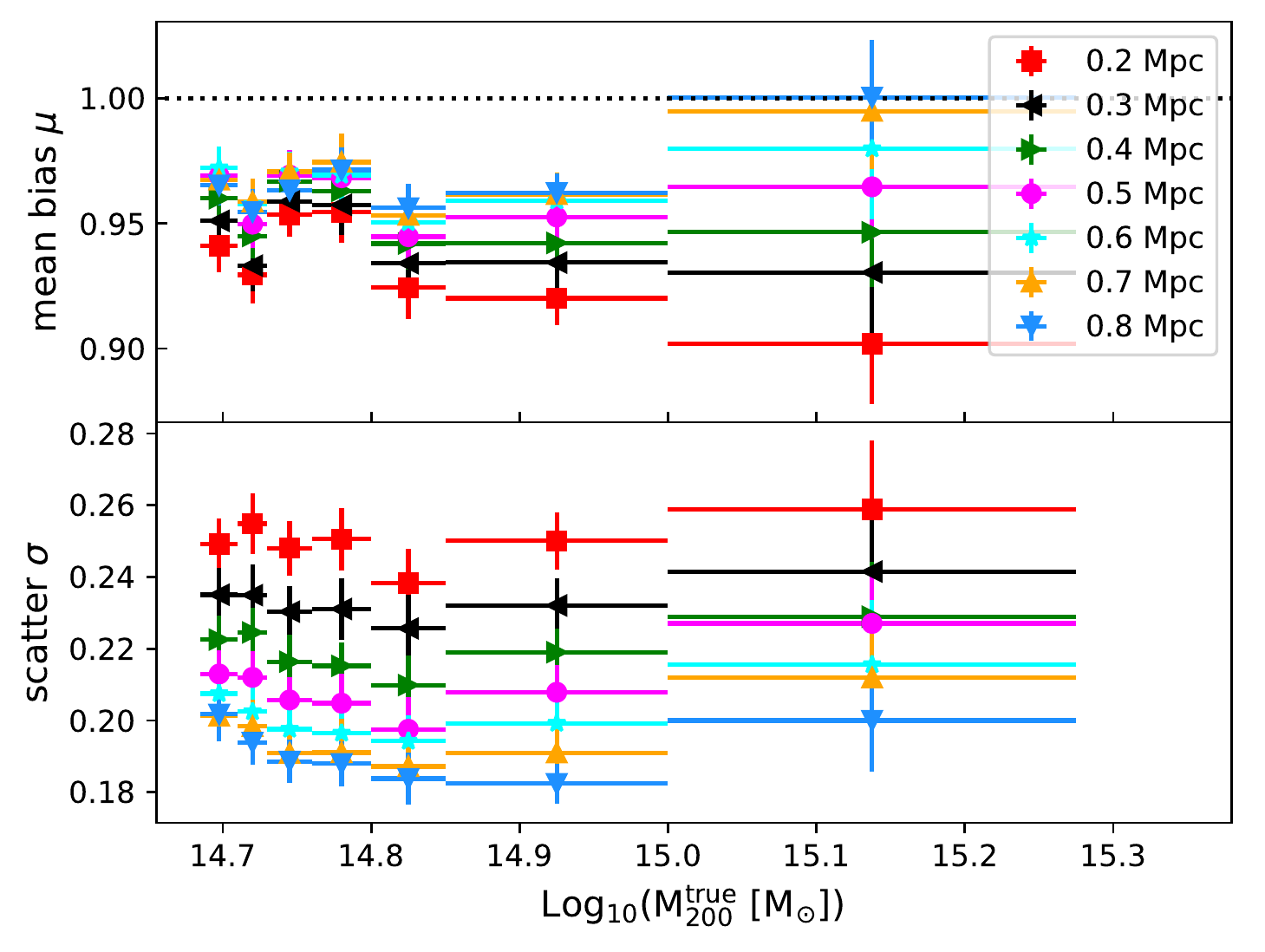} 
\includegraphics[width=0.48\textwidth,clip=True,trim={0 0 -10 5}]{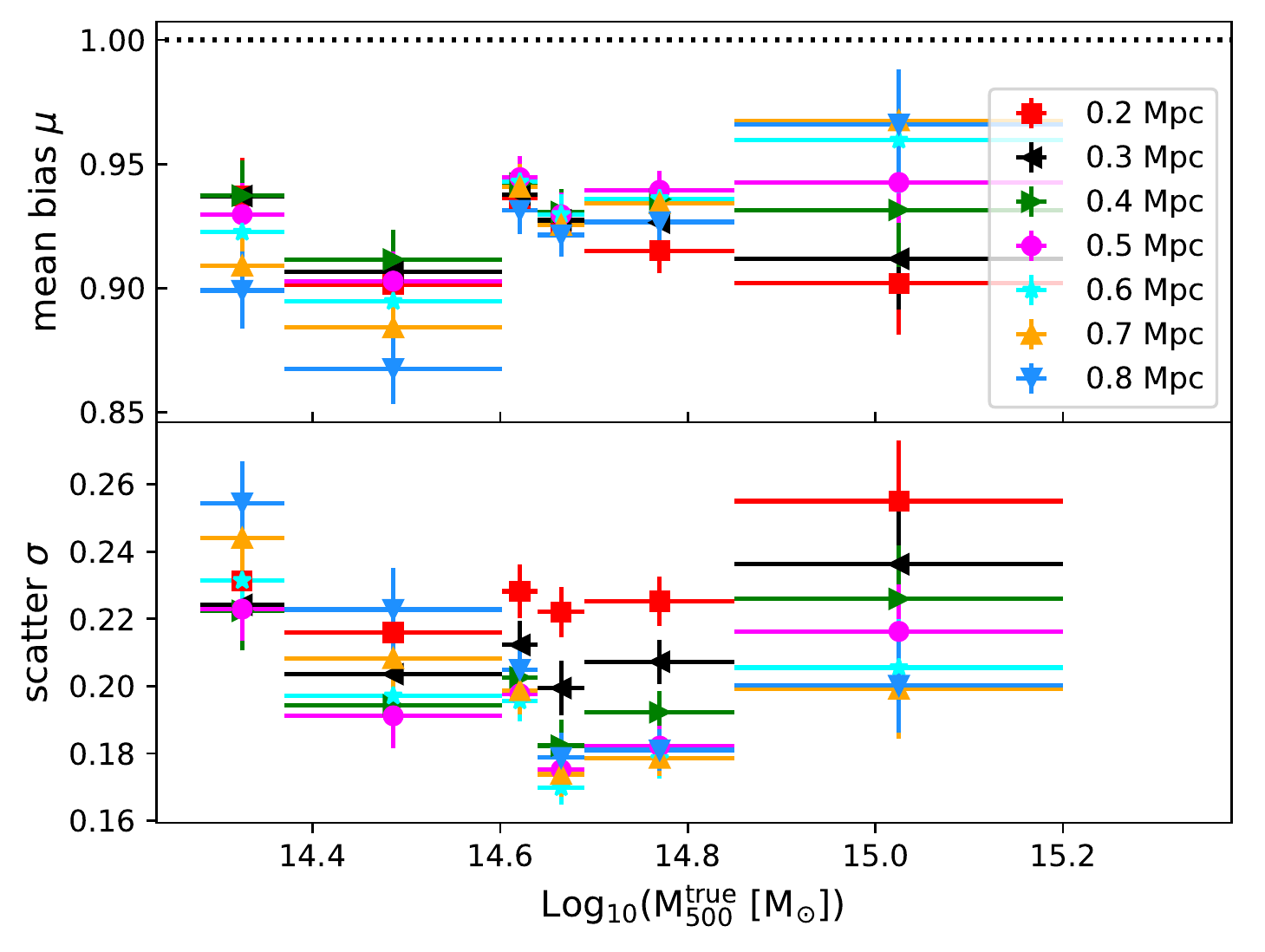}
\caption{Bias distributions as a function of mass at $z=1$ (MXXL simulation) for different radial fit ranges. The legend in each panel indicates the inner radius, while the outer radius was kept constant at the default value of $3.5$ Mpc. Left: $\mth$. Right: $\mfh$.}
\label{fig:biasdep:innerradius}
\end{figure*}

Because of discrepancies between simulated halos and the NFW profile close to the halo center, we expect that the minimum radius limiting the mass fit from reduced shear will have a considerable impact on the mass bias. We vary the inner radius $r_{\rm{min}}$ in the range 0.2$-$0.8 Mpc while keeping the outer radius $r_{\rm{max}}$ constant at the fiducial value. The results are shown in Fig.~\ref{fig:biasdep:innerradius}. 
As expected, the mean bias increases with decreasing $r_{\rm{min}}$, with a simultaneous increase in the scatter. The trend is present in the full range of masses, though the picture is not completely clear for $\mfh$ at low mass. 

\subsubsection{Concentration$-$mass relation}
\label{sec:res:biasdep:mc}

\begin{figure*}	
\includegraphics[width=0.48\textwidth,clip=True,trim={0 0 -10 5}]{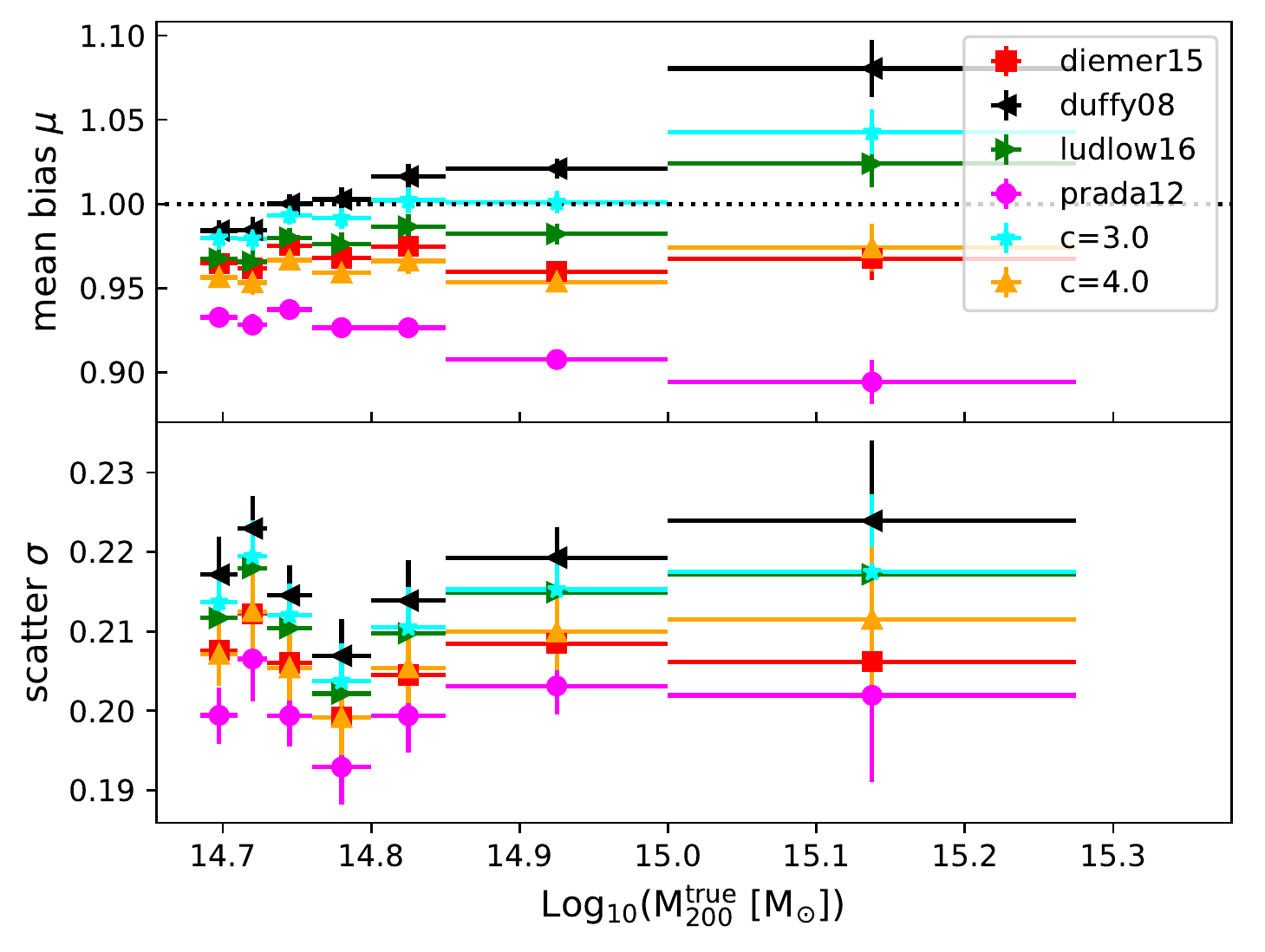} 
\includegraphics[width=0.48\textwidth,clip=True,trim={0 0 -10 5}]{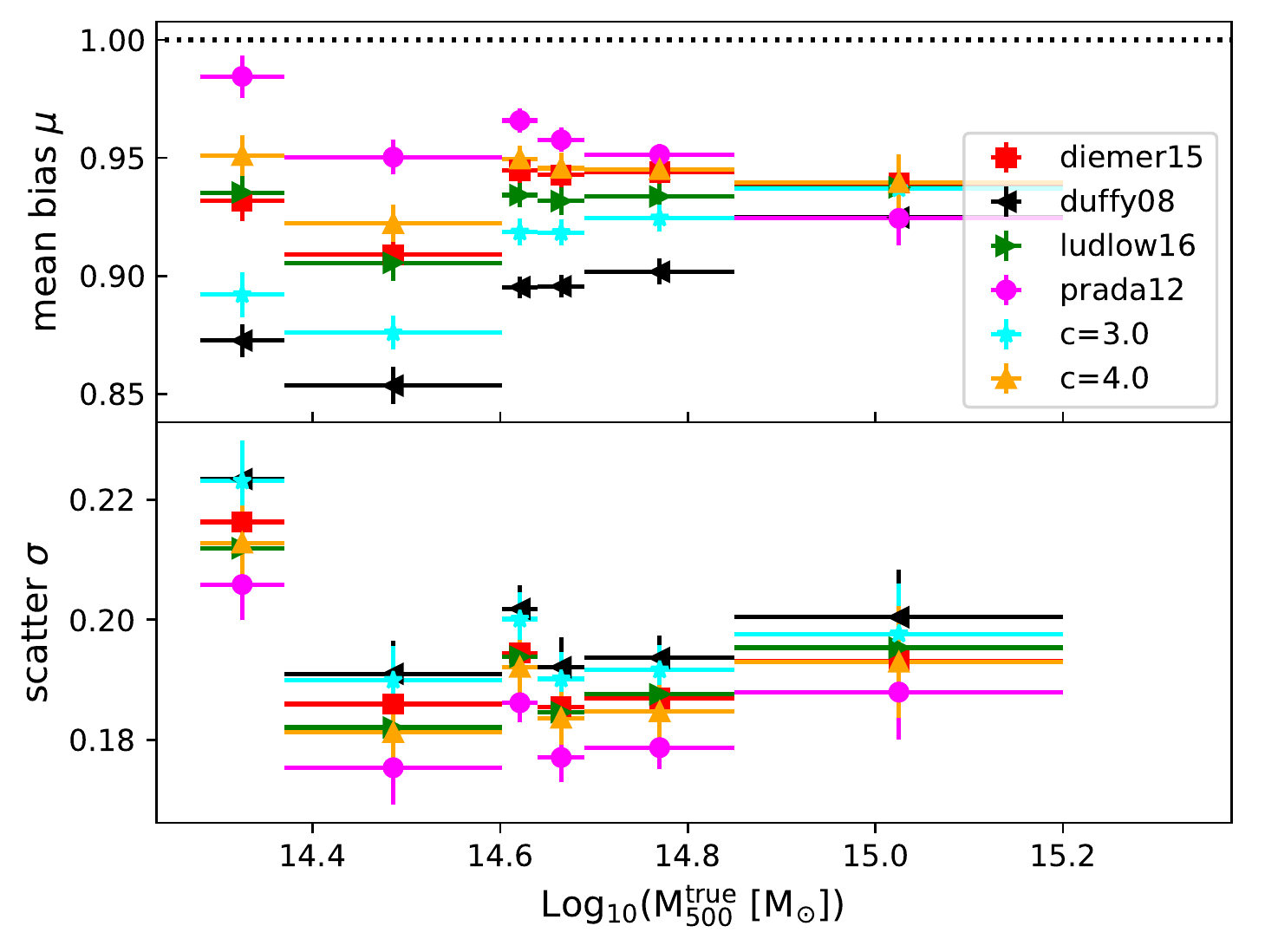} 
\caption{Bias distributions as a function of mass at $z=1$ (MXXL simulation) for different concentration$-$mass relations, and for two cases of constant concentration $c = c_{200}$ in the NFW model. The radial fit range of the reduced shear is $0.5-3.5$ Mpc. Left: $\mth$. Right: $\mfh$. As in the fiducial setup, we have used the updated parameters from \citet{2019ApJ...871..168D} for the \citet{2015ApJ...799..108D} model. }
\label{fig:biasdep:mcrel}
\end{figure*}

Because we use a concentration$-$mass relation to avoid the degeneracy in the NFW model, the bias distribution will also depend on the choice of such a relation. In Fig.~\ref{fig:biasdep:mcrel} we compare the results from some of the concentration$-$mass relations mentioned in \sect~\ref{sec:method:nfw}, specifically those of \cite{2008MNRAS.390L..64D,2012MNRAS.423.3018P,2015ApJ...799..108D,2016MNRAS.460.1214L}. In addition, we also consider two cases with constant $c_{200}$. 

While this comparison is by no means exhaustive, it underlines the importance of considering this aspect, as the mean bias for $\mth$, for example, varies between $0.9$ and $1.1$ at high mass depending on the relation used. Notably, the differences between the various relations considered here are largest at high mass when $\mth$ is considered, while the discrepancies are larger at the low masses when considering $\mfh$.


\subsection{Convergence SNR miscentring distributions}
\label{sec:res:kappamis}

In this subsection we describe the miscentring distributions resulting from using the peak of the reconstructed convergence signal-to-noise ratio (SNR) image as the estimator for the halo center. Naturally, such distributions are critically dependent on the noise level. We quantify this dependence in terms of the peak SNR. 

\subsubsection{Convergence SNR estimation}
\label{sec:res:kappamis:snr}

Comparing the three randomization methods described in \sect~\ref{sec:meth:miscentring} (random phase, random position, bootstrapping) to the reference (independent noise realizations), we find that all methods yield SNR values very close to the reference. On average, the bootstrapping method overestimates the SNR by around $5\pm0.3\%$, randomization of ellipticity phases underestimates the SNR by $0.7\pm0.3\%$, and the randomization of positions is consistent with the reference ($+0.3\pm0.2\%$). For the following results, we rely on the phase randomization technique, as it can be used also in case of a variable galaxy density in the field. 

\subsubsection{Convergence peak miscentring distributions}
\label{sec:res:kappamis:distr}

\begin{figure}
\centering
\includegraphics[width=\columnwidth,clip=True,trim={0 0 0 0}]{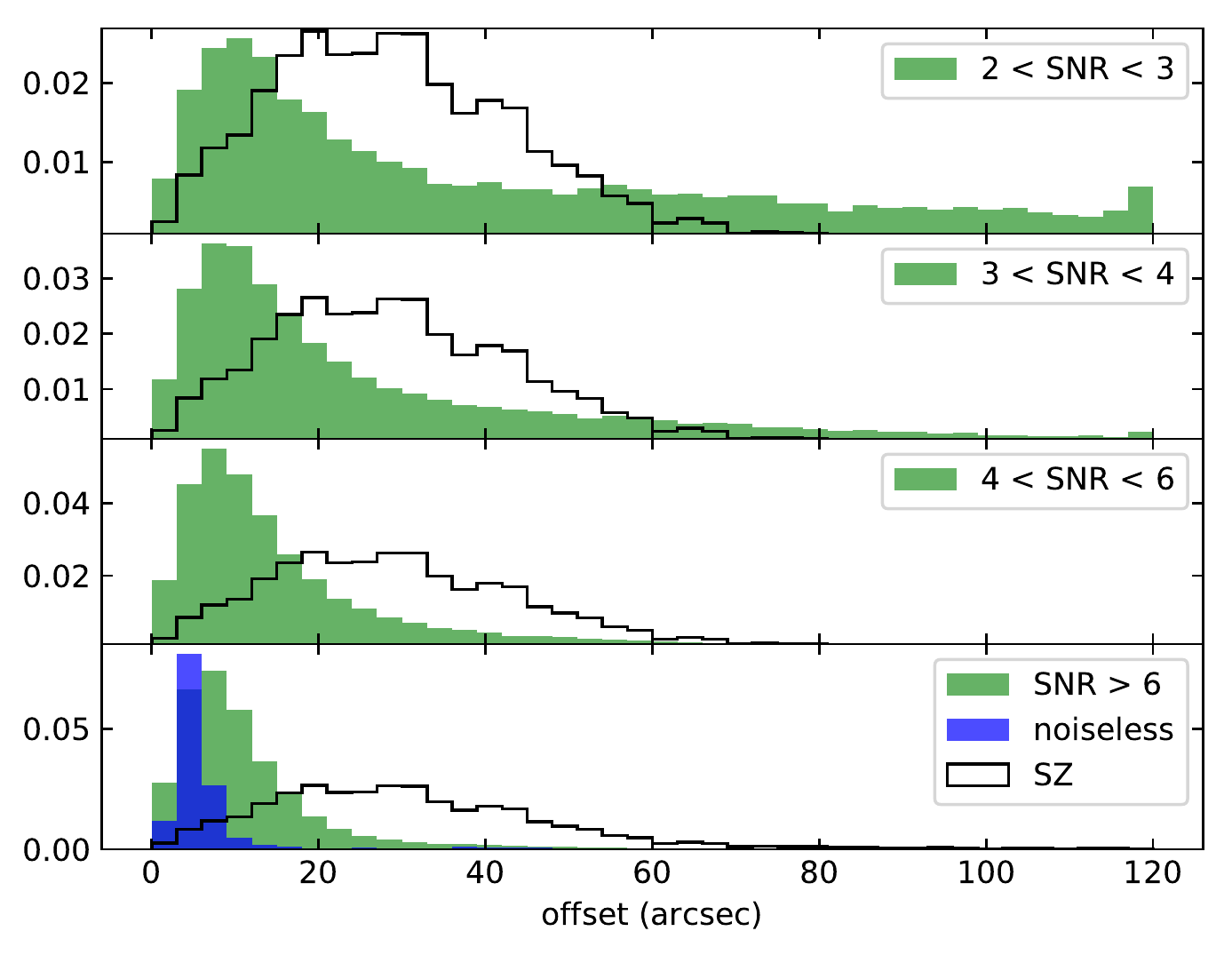}
\caption{\label{fig:offsets.kappa.sz} Miscentring distributions derived from using the convergence SNR peak in 500 simulations at $z=1$, binned by peak SNR (green). In the bottom panel, we show the miscentring distribution in the ideal case of noiseless simulations (blue), where the distribution arises purely by projection effects.  In all panels, the averaged SPT-SZ miscentring distribution described in \sect~\ref{sec:res:miscenter} is indicated (black). } 
\end{figure}

Binning the fields by peak SNR and measuring the offsets of each peak from the nominal position (the center of the halo as defined by the most bound particle in the simulation) yields empirical miscentring distributions for convergence-derived centroids. We limit the offset to a maximum of two arcminutes at $z=1$. The resulting miscentring distributions are
shown in Fig.~\ref{fig:offsets.kappa.sz}. We compare the results to the SZ miscentring distribution described in \sect~\ref{sec:res:miscenter}, averaged over empirical parameters.

For comparison, we also show the offset distribution of a virtually noiseless simulation (using each pixel in the simulation of reduced shear as a source galaxy, and with no added shape noise). The latter distribution, peaking at around 5 arcseconds, arises due to projection effects (the simulated halos are not spherically symmetric). Even at low SNR (two to three at the convergence peak) the miscentring distribution peaks at a lower offset than the corresponding SZ derived distribution; however, it also shows a wide tail with a significant fraction extending well beyond the maximum SZ derived offsets. For SNR greater than 4, the convergence peak is clearly preferred in terms of positional accuracy.   

\begin{figure}
\centering
\includegraphics[width=0.95\columnwidth,clip=True,trim={0 0 0 0}]{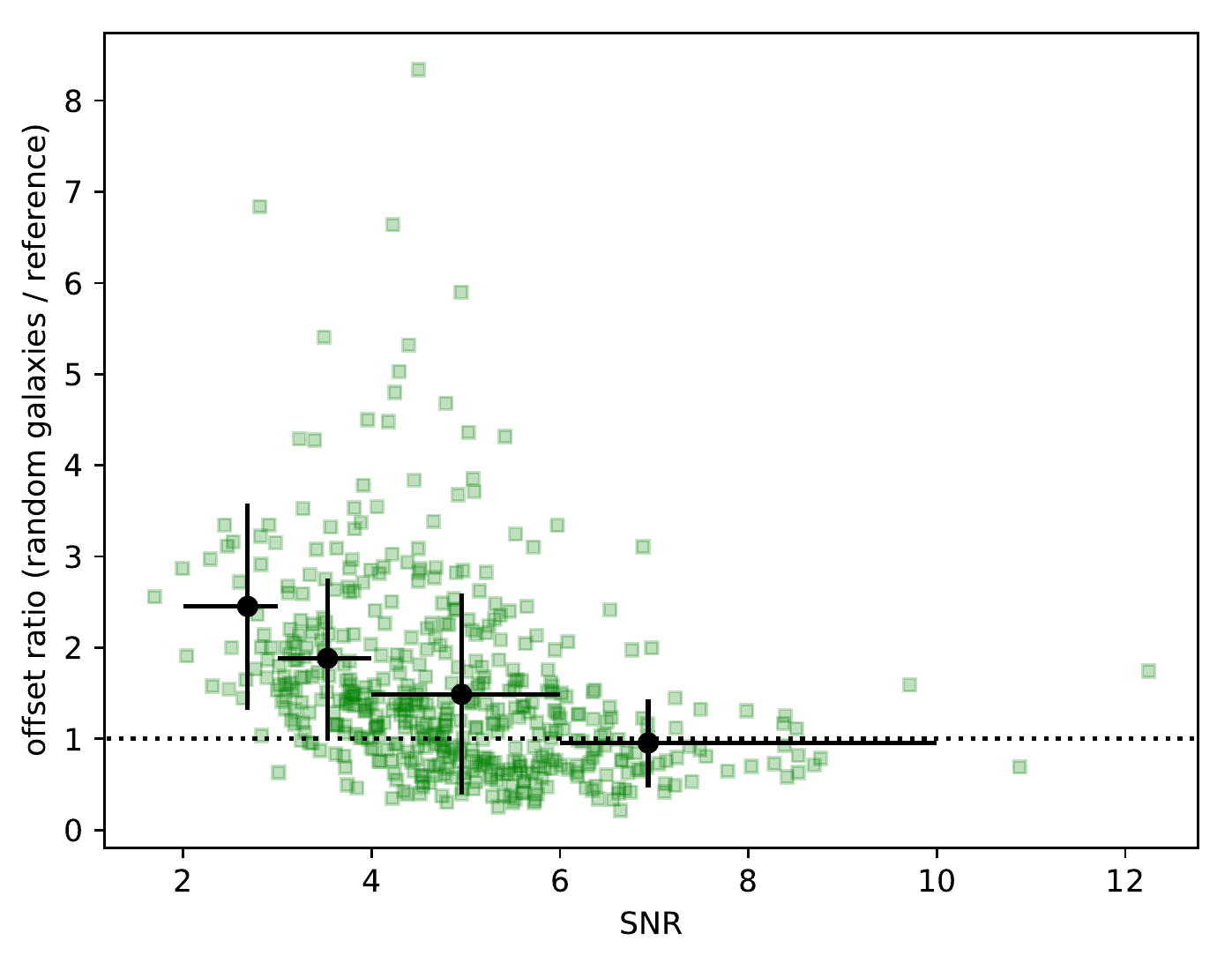}
\caption{\label{fig:random.gal.offsets.bias} For individual simulated clusters (green squares), we show the ratio of derived (from bootstrapping) to expected (from the reference sample) average positional offsets of the convergence peak from the mean position of the latter across all 
bootstraps/realizations, as a function of SNR. The black points with error bars indicate averages in SNR bins. Vertical error bars indicate standard deviations in bins, while horizontal error bars show the binning by SNR. } 
\end{figure}

\begin{figure}
\includegraphics[width=0.5\textwidth,clip=True,trim={15 0 0 -10}]{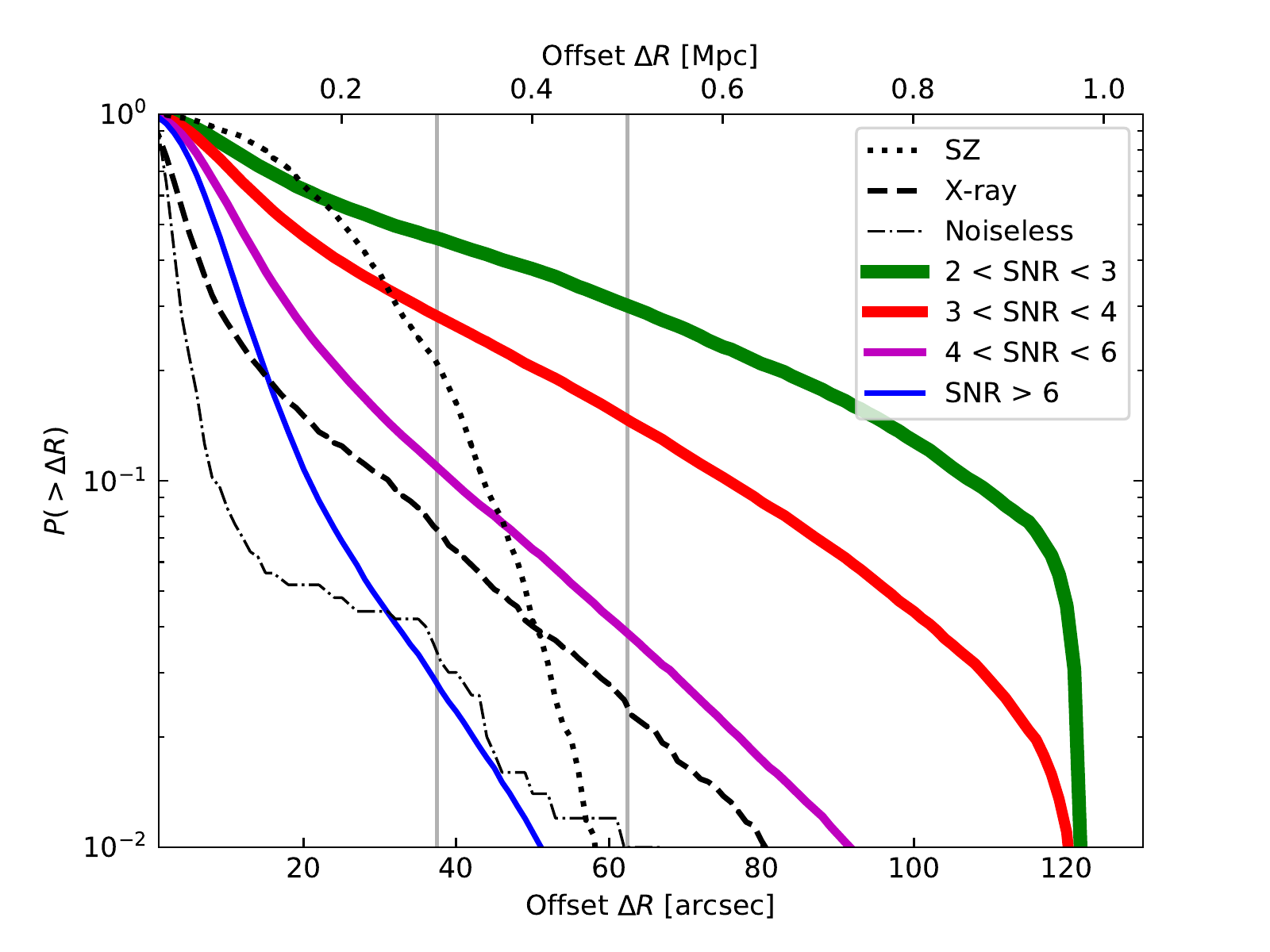} \\
\caption{Integrated miscentring distributions used and derived in this work. The plot shows the probability of the miscentring offset being greater than a radial coordinate (evaluated here at redshift $z=1.0$). The transformation between physical and angular coordinates is valid for $z=1$. For comparison, we also show the two physical radii at which we cut off the inner part of the shear profile in this work. Distributions from convergence peaks (derived in \sect~\ref{sec:res:kappamis:distr}) are shown as solid lines for different ranges of signal-to-noise ratios. The sharp cutoffs at low signal-to-noise are artificial; the search for a peak was limited to a radius of two arcminutes.}
\label{fig:miscdisr.xszo}
\end{figure}

\subsubsection{Positional uncertainty in convergence images}

The position of the peak in the convergence SNR image has an associated uncertainty with respect to the position of the bottom of the gravitational potential. Bootstrapping by selecting random entries from the background galaxy catalog yields an estimate of this uncertainty. Here, we test the robustness of this estimate by comparing it to the spread in position from the reference simulations, where each SNR image comes from an independent noise realization. For simplicity, we characterize the positional distribution by the average offset from the nominal position across all noise realizations of a target field. 

For each target field, we then compute the ratio of averages (from bootstrapping versus from independent noise realizations). In Fig.~\ref{fig:random.gal.offsets.bias}, we show this ratio as a function of peak SNR. While there is a large spread in the ratio, we see a clear tendency of overestimating the centroid uncertainty at low SNR. In fields with a signal-to-noise ratio greater than 6 at the peak of the convergence, we find that this bias vanishes.

\subsection{Including miscentring in the bias estimation}
\label{sec:res:miscenter}

\begin{figure*}
\centering
\includegraphics[width=0.98\textwidth,clip=True,trim={70 10 80 30}]{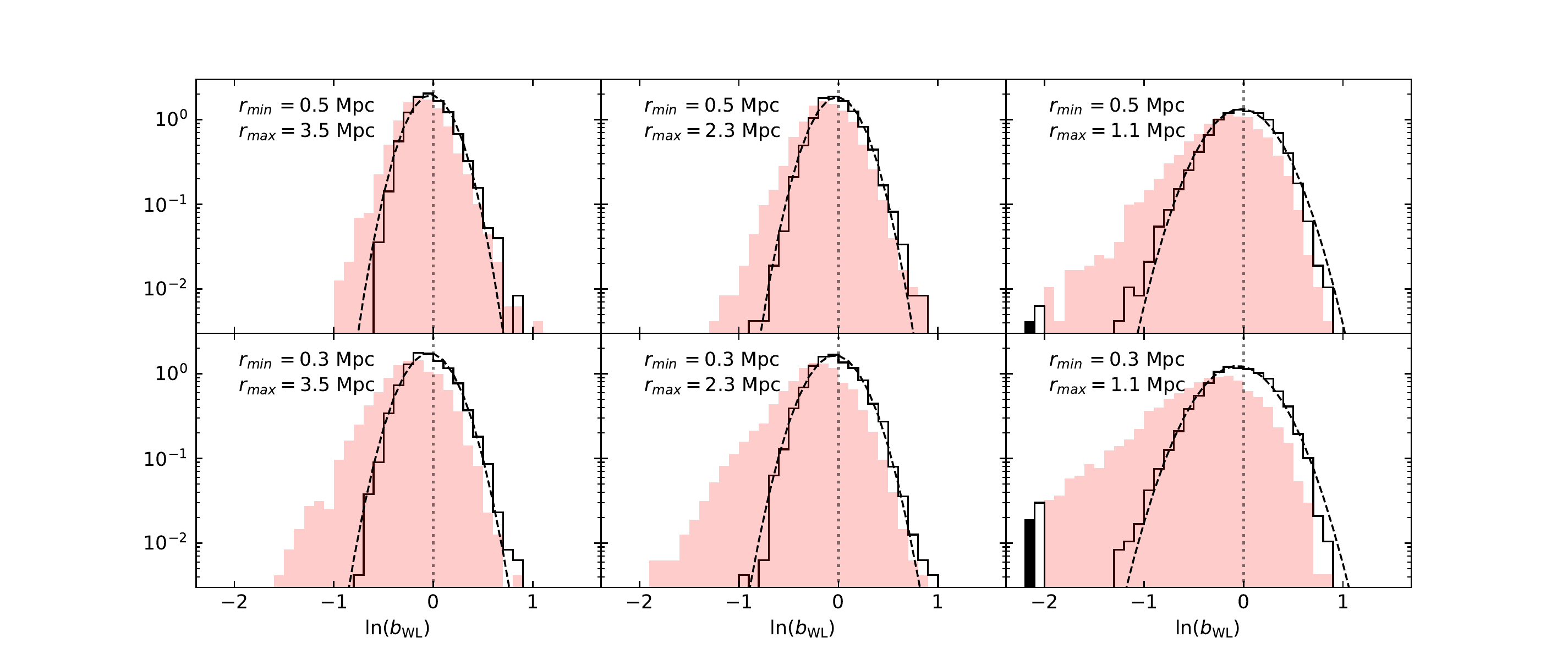}
\includegraphics[width=0.98\textwidth,clip=True,trim={70 10 80 30}]{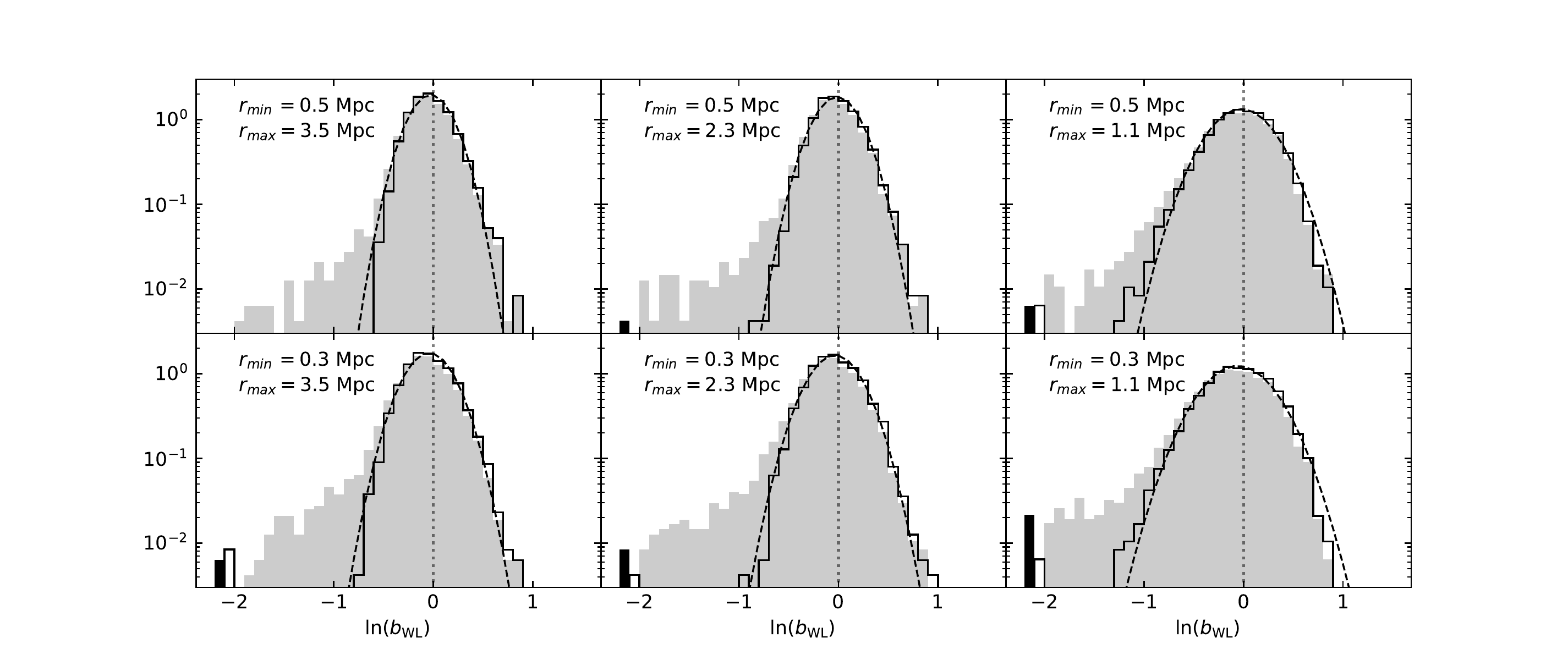}
\caption{\label{fig:biashisto:m200} Weak lensing mass bias distributions (histograms) from noiseless simulations in the presence of miscentring, from the MXXL simulation at $z=1$ and for $\mth$. Top: SZE miscentring. Bottom: X-ray miscentring. Each panel represents a different radial range for the mass fit from reduced shear. Line histograms represent perfectly centered halos, with dashed curves representing corresponding log-normal distributions with the same mean and sample variance. Filled histograms indicate bias distributions including miscentring. For ease of viewing, all histograms were cut at $\ln(\mu) = -2$, and all lower values of $\mu$ are indicated by a light outlined bar (positive $\mu$) and a dark filled bar (negative $\mu$, corresponding to fitted negative masses). The $y$-axis scale is arbitrary.} 
\end{figure*}

\begin{figure*}
\centering
\includegraphics[width=0.98\textwidth,clip=True,trim={70 10 80 30}]{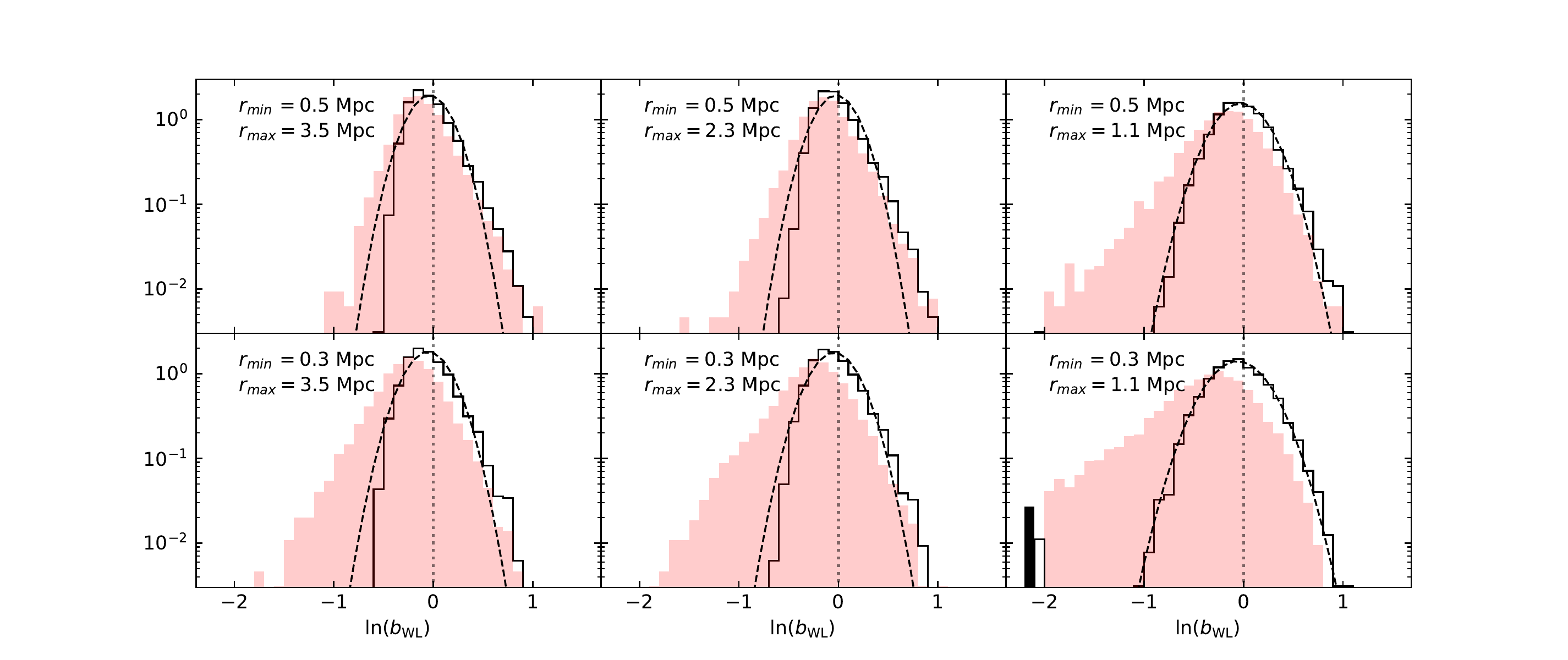} 
\includegraphics[width=0.98\textwidth,clip=True,trim={70 10 80 30}]{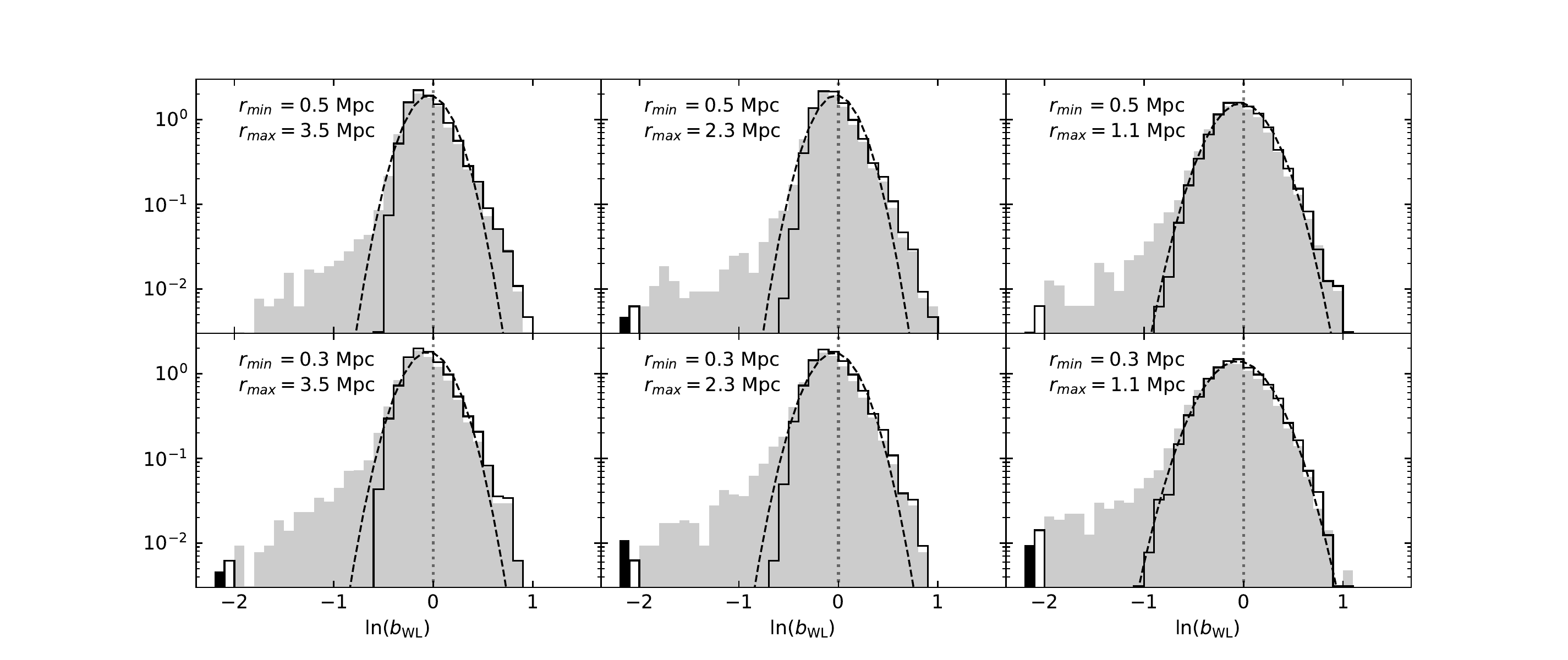}
\caption{\label{fig:biashisto:m500} As Fig.~\ref{fig:biashisto:m200}, but for $\mfh$.}
\end{figure*}

Thus far, we have considered only perfectly centered halos for the determination of the WL mass bias distributions, in the sense that we have used the position of the most bound particle in the simulation for the construction of each reduced shear profile. We now turn our attention to how the bias distribution changes when miscentring is directly included in the analysis. As in the previous subsections, we fit azimuthally symmetric NFW profiles to the reduced shear images; however, we center the profiles on a coordinate randomly chosen from one of several miscentring distributions. We do not consider the approach of leaving the center coordinate free to vary in the fit. 

In addition to the miscentring distribution from convergence centers (previous subsection), we use two specific miscentring distributions derived for typical SZE observations and for typical X-ray center determinations. The distributions were derived based on the Magneticum Pathfinder Simulation \citep{2016MNRAS.463.1797D}, and are described in detail by \cite{2020arXiv200907591S}. We summarize the most important points here. To replicate the observing conditions of the SPT$-$SZ survey, thermal SZE light-cones were
built, from which mock SPT observations were extracted. In each mock, contributions from primary CMB anisotropies, the SPT beam
and transfer function \citep{2011ApJ...743...90S}, and instrumental noise were accounted for. Cluster candidates were identified with the approach adopted for SPT clusters \citep[e.g.][]{2009ApJ...701...32S}. Different cluster core
sizes $\theta_{\mathrm{c}}$ were adopted, in line with the SPT data analysis. While there is a dependence on this parameter for the miscentring distribution, this dependence is much weaker than the difference between the SZE and convergence miscentring distributions, and is neglected in this work. The resulting sample of SPT-like selected clusters was used to characterize both the SZE and X-ray miscentring distributions. Cut-outs of
X-ray surface brightness maps were produced at the point of the deepest potential of each halo. The X-ray miscentring distribution was then derived as the distribution
of the projected offsets between the peak of the X-ray surface brightness maps and the
position of the deepest potential in the halo.

We also investigate the effects of using the peak of the convergence signal-to-noise ratio (SNR). The associated miscentring distributions are derived as a function of SNR in \sect~\ref{sec:res:kappamis:distr}. 
Our simulated weak lensing data have the same noise properties for all clusters in terms of source density, shape noise, average lensing efficiency and field size. In order to compute the SNR images we can therefore re-use noise images, reducing processing time. To this end, we derive a mean noise image from 100 randomly chosen halo fields, and apply this average to all fields to construct SNR images.
We center each shear profile on the convergence SNR peak inside a search radius of 2 arcminutes, while for the SZE and X-ray miscentring modes we generate random instances of the corresponding distributions.  

The characteristics of the miscentring distributions are shown in Fig.~\ref{fig:miscdisr.xszo}. While the X-ray distribution is tighter than the SZE distribution overall, it has a much wider tail. The miscentring distributions from convergence centering are inferior to both X-ray and SZE miscentring at low SNR. 

It is natural to expect that masses will be additionally biased low when halo centers are randomly offset using a miscentring distribution, while the scatter of the mass bias distribution is expected to increase due to the additional scatter introduced by the miscentring. In the case of using the convergence peak for the centring, however, due to the correlation of convergence and reduced shear, the situation is not as clear, and at least from the inner part of the shear profile one would expect masses to be overestimated as shear and convergence are not independent. 

As we have seen, the inner and outer limiting radii of the shear profile fit play a prominent role in the mass bias. With miscentring, naturally, we expect this effect to be amplified.

We investigate the effects of the various modes of miscentring by generating shear and convergence from the MXXL simulations at $z=1$, using different noise levels and different radial ranges for the mass fits. 

In Figs.~\ref{fig:biashisto:m200} and \ref{fig:biashisto:m500} we show how the WL mass bias distribution from noiseless mass fits are affected by miscentring in different radial fit ranges. Here we show the distributions for all masses, excluding the first and last mass bin. Clearly none of the distributions are truly log-normal. The non-miscentered distributions are in all cases close enough to log-normal that the discrepancy in the mean and the median of $\ln b$ is at the sub-percent level. The bias distributions from miscentered halos generally show deviations on the order of several per cent. In the mass fitting we have included ``negative'' masses by allowing a negative sign in the normalization of the density profile (keeping $c_{\Delta}$ positive). These occur in a small percentage of the miscentered halos, and must necessarily be excluded when comparing means and medians in log-space. Equivalently, one may restrict the analysis to non-negative masses, yielding a bi-modal distribution with a sharp peak at vanishing mass.

We next investigate the noise dependence of the mass bias parameters in the presence of the various types of miscentring. While we established that the mass bias distribution is essentially independent of noise for perfectly centered halos (\sect~\ref{sec:res:noise_indep}), the same is not necessarily the case when we fit a log-normal distribution to an underlying distribution that is in fact not log-normal. 

\begin{figure*}
\centering
\includegraphics[width=0.79\textwidth,clip=True,trim={0 37 0 0}]{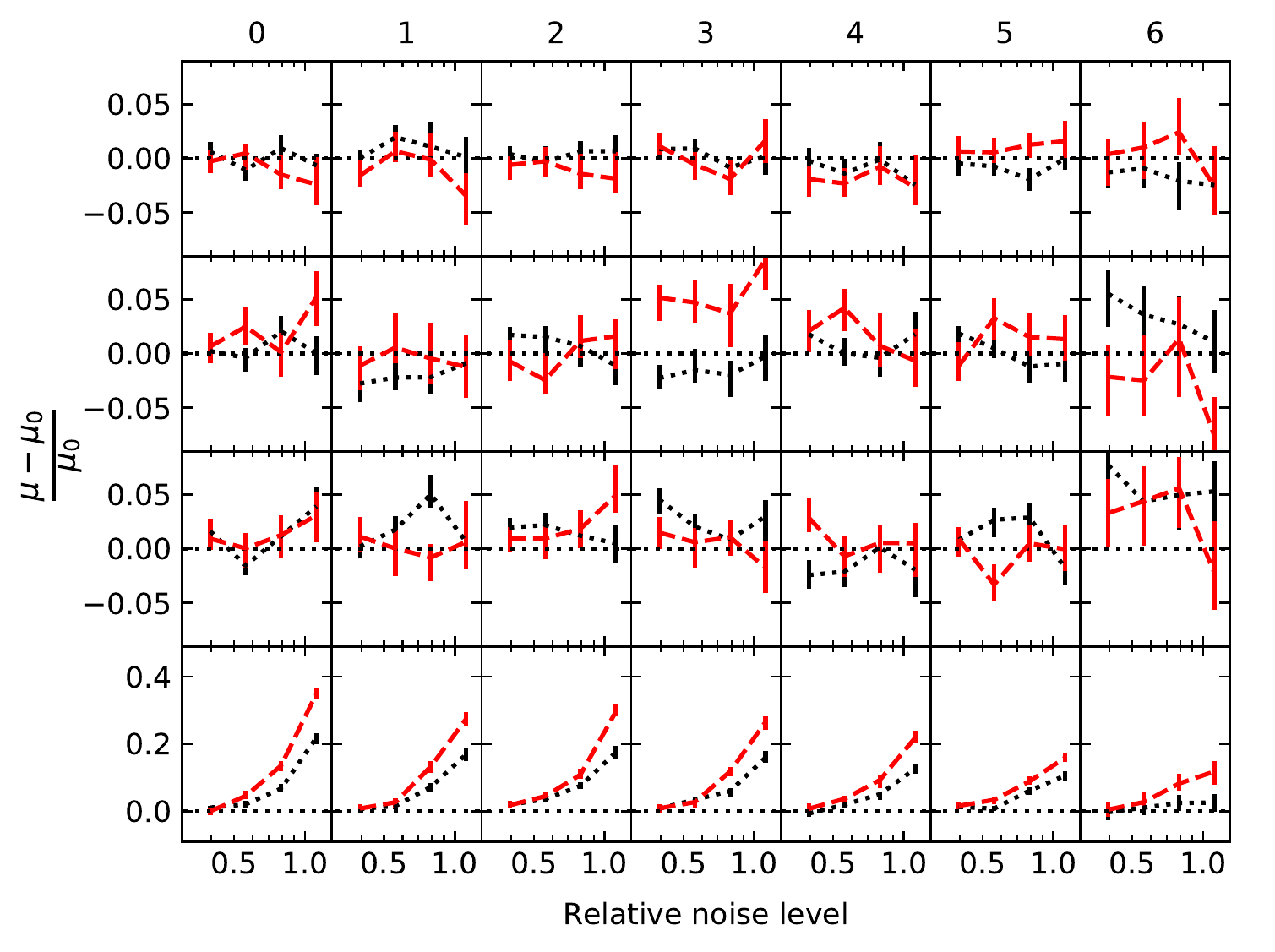}\\
\includegraphics[width=0.79\textwidth,clip=True,trim={-2 0 1 17}]{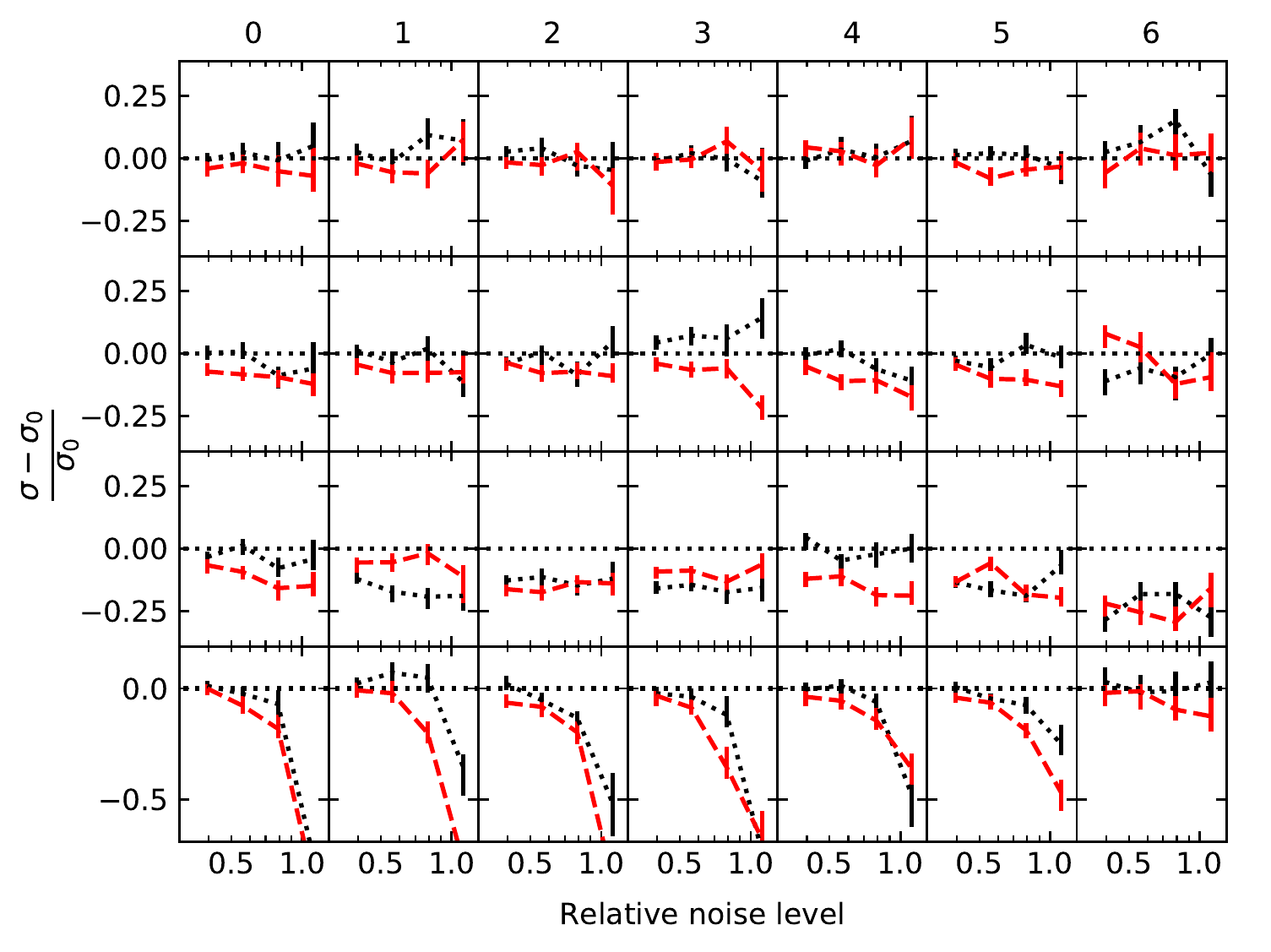}
\caption{\label{fig:miscentr.relative.200} Weak lensing mass bias mean (top) and scatter (bottom), relative to the corresponding values derived from noiseless realizations, as a function of the relative noise level, from the $z=1$ slice of the MXXL simulation and for $\mth$. The reduced shear was fitted in two different radial ranges: $0.5 - 2.3$ Mpc (black dotted lines) and $0.5 - 1.1$ Mpc (red dashed lines). The noise level is normalized to one at shape noise 0.25 and a background galaxy density of 10 arcmin$^{-2}$. Bin numbers are indicated at the top of each sub-figure, with mass increasing from left to right. In each sub-figure, the top row represents perfectly centered halos, the second and third rows were realized using SZE and X-ray miscentring distributions, respectively, and the bottom row was realized using the convergence SNR peak in each halo realization for centering.} 
\end{figure*}

\begin{figure*}
\centering
\includegraphics[width=0.79\textwidth,clip=True,trim={0 37 0 0}]{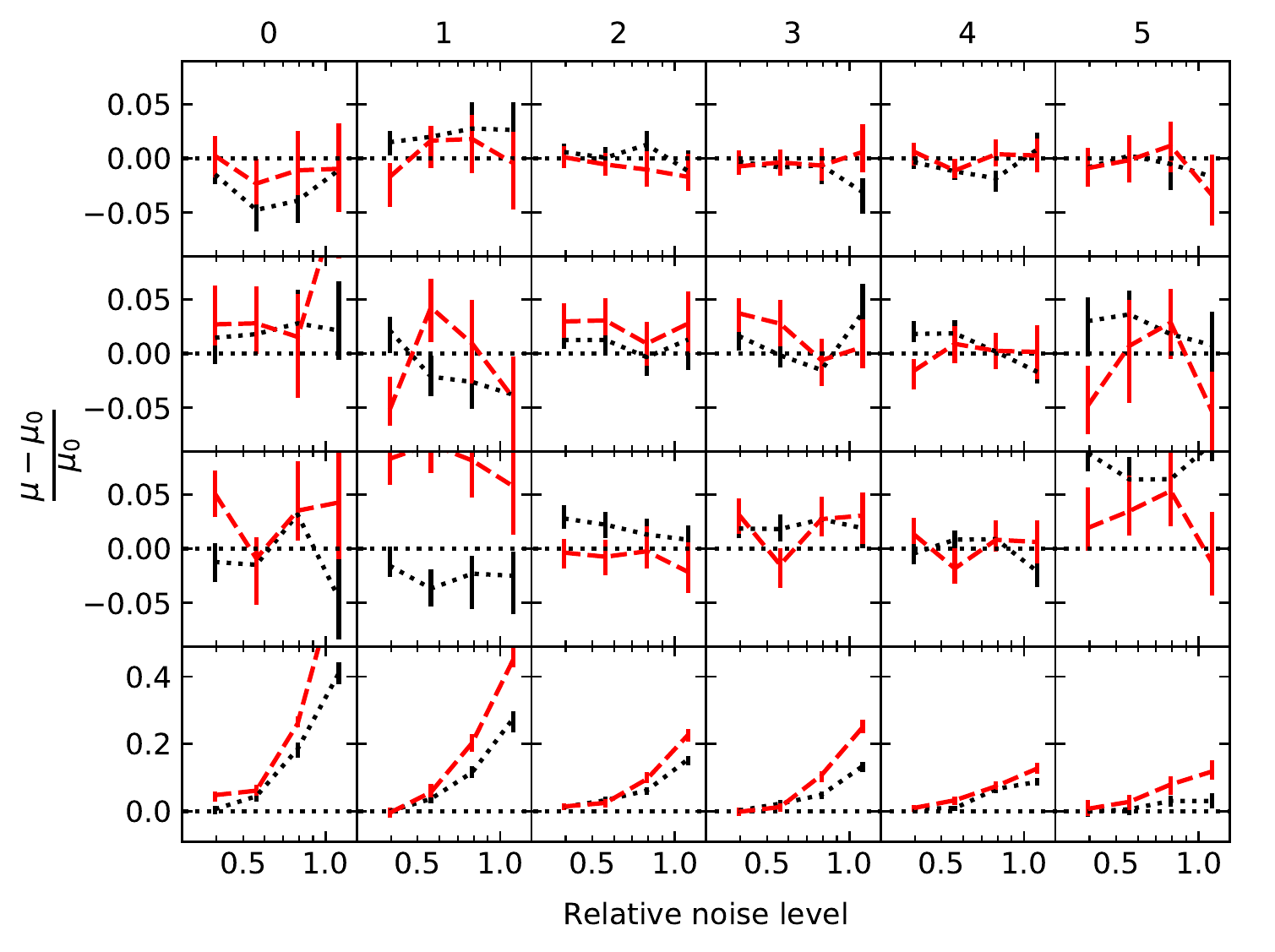}\\
\includegraphics[width=0.79\textwidth,clip=True,trim={-2 0 1 17}]{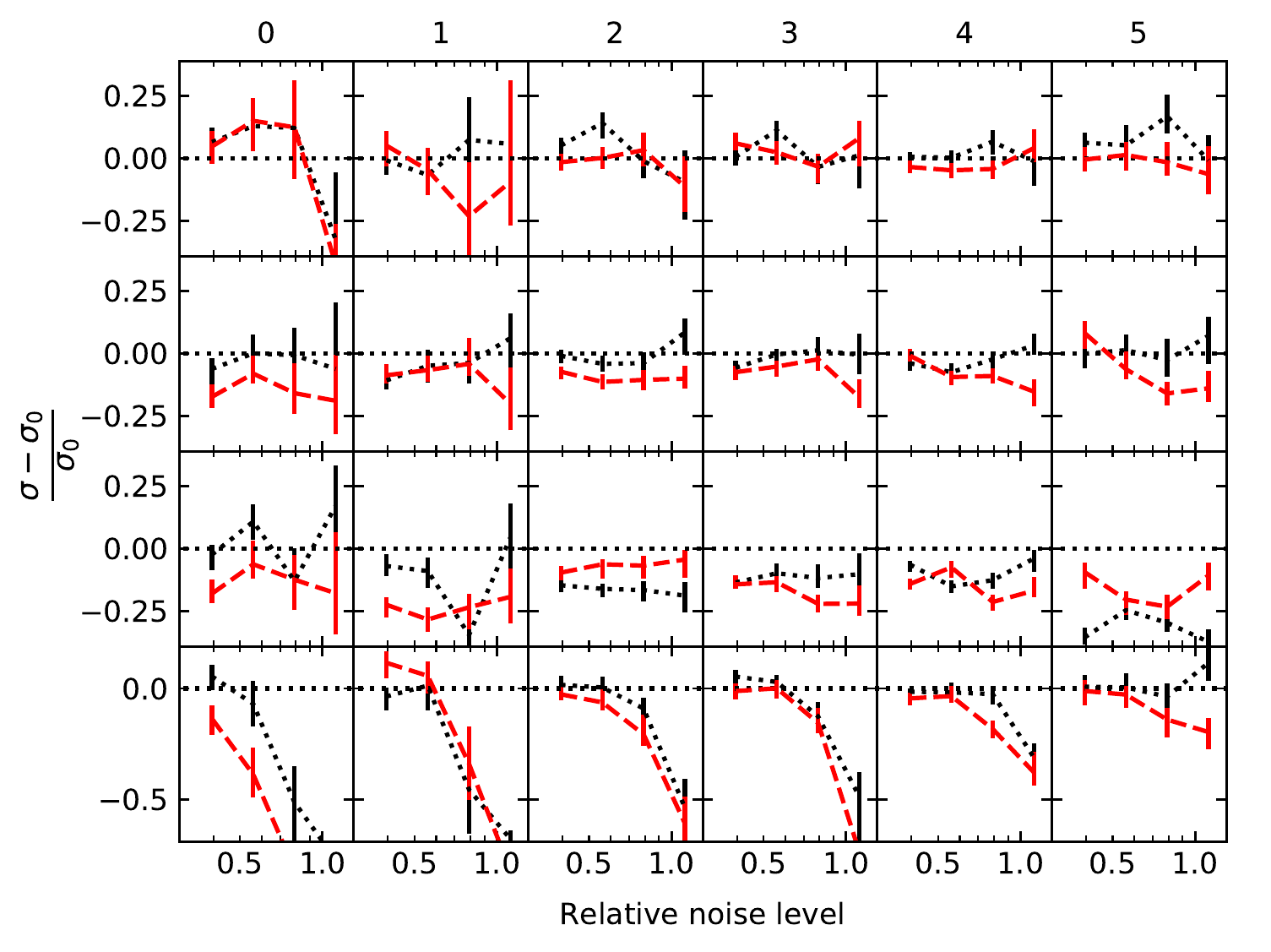}
\caption{\label{fig:miscentr.relative.500} As Fig.~\ref{fig:miscentr.relative.200}, but for $\mfh$.} 
\end{figure*}

At redshift 1, we compute the mass bias for two radial ranges of reduced shear, namely $0.5 - 2.3$ Mpc and $0.5 - 1.1$ Mpc. In Figs.~\ref{fig:miscentr.relative.200} and \ref{fig:miscentr.relative.500}, we show the bias parameters $\mu$ and $\sigma$, relative to their noiseless counterparts, as a function of the relative noise level. As expected, with no miscentring (top rows in each panel) the results are consistent with no mass dependence, since the distributions are close to log-normal. The X-ray and SZE miscentring distributions introduce up to $\sim$6\% and $\sim$20\% discrepancy in the mean bias and bias scatter, respectively.  Especially pronounced is a systematic decrease in the scatter (around $10\%$ at intermediate mass) in the X-ray miscentring case.  

As we might expect, the mean bias increases sharply with increasing noise in the convergence SNR peak miscentring scenario, as the miscentring distribution is dependent on the noise level (which is not the case for the SZE and X-ray miscentring distributions). Simultaneously, the scatter decreases sharply.
These effects decrease with increasing mass. This is expected, since the signal-to-noise ratio of the kappa images also increases with mass, yielding a center proxy closer to the true center (the bottom of the gravitational potential). With increasing noise, the convergence peak will shift in a direction which increases the measured (noise-boosted) tangential reduced shear. Thus we naturally expect a measurement that is biased high.

\section{Discussion}
\label{sec:discussion}

\subsection{Bias level and scatter}

Previous studies have generally found that WL masses are biased low by $\sim 5-10\%$ on average, with the bias decreasing with increasing mass \citep{2011ApJ...740...25B,2011MNRAS.414.1851O,2012MNRAS.421.1073B, 2017MNRAS.465.3361H,2018MNRAS.479..890L}. 
We find some results consistent with this tendency; however, in general we find large differences depending on the radial range and the concentration$-$mass relation used. For our fiducial radial range of $0.5-3.5$ Mpc, we found results consistent with less negative bias at high mass for some of the concentration$-$mass relations in regards to both $\mfh$ and $\mth$.
The choice of concentration (whether constant or bound to the mass) will lead to different slopes in the mass dependence contingent upon the choice of overdensity, which is also obvious from our results (see Fig.~\ref{fig:biasdep:mcrel}). 
The level of scatter is less sensitive to the choice of concentration$-$mass relation and the radial range, changing by at most two to three percent (with respect to a mean bias of 1) over the range of masses considered. 

A detailed comparison with previous publications in terms of the bias level and scatter is in most cases neither possible nor appropriate, considering the many differences in the experimental setup. In particular, the bias distribution is critically dependent upon the radial range considered, and upon the choice of concentration in the NFW density profile. 
While we have exclusively considered NFW mass profiles in this work, it would be expected that a different mass model would similarly result in bias distributions different from those presented here. Conversely, given a set of observations it is straightforward to model the mass bias distribution by applying the same model to both data and simulations. 

\cite{2011MNRAS.414.1851O} and \cite{2018MNRAS.479..890L} did not use concentration$-$mass relations, making it difficult to investigate a noise dependence in the bias distribution to high accuracy, especially at low mass, requiring very large numbers of halos in the simulation. While Lee et al. did find a noise dependence under the assumption of a log-normal bias distribution, we speculate that the origin of this behavior may actually be due to deviations from the log-normal distribution. Another possible explanation would be that at high noise levels, the tails of the probability distributions for masses of individual halos are not sufficiently sampled. Such effects could be amplified when fitting for mass and concentration, as these can be highly degenerate. However, it cannot be ruled out that non-linear aspects of the fitting introduce a real noise dependence in the bias. The existence of such a dependence would make the modeling of the bias on observations rather complicated. 

We perform a direct comparison with the results of (\citealt{2021arXiv210316212G}, henceforth G21), using a near-identical setup. Using a constant concentration of $c_{200} = 3.5$, a fit range of $0.5-2.58$ Mpc/$h$ at redshift $z=0.24$ (from hydro-dynamical simulations) and a minimum mass of $M_{200} = 2.85 \times 10^{14} M_{\odot} / h$, G21 found $\ln \mu = -0.033$ and $\sigma = 0.19$, with virtually no mass dependence. Reproducing the relevant parameters in our analysis at a similar redshift using MXXL snapshot 54 (BK11 snapshot 141), combining all our mass bins above the minimum mass, we find $\ln \mu = -0.028 \pm 0.005$ ($-0.016 \pm 0.012$ from BK11), in agreement with G21, and a slightly higher scatter $\sigma = 0.22 \pm 0.005$ ($0.24 \pm 0.01$ from BK11). Using mass bins, we see no significant change of $\mu$ with mass, although $\sigma$ shows a slight increasing trend with increasing mass. At the lowest masses our results from MXXL are consistent with the results of G21. We speculate that the small differences seen in the scatter arise due to differences between the n-body simulations used here and the hydro-dynamical simulations used by G21.

\subsection{Noise independence of the mass bias}

One main result of our paper is that under the assumption of a log-normal bias distribution and under the use of a concentration$-$mass relation, the weak lensing mass bias distribution can be modeled to within a few per cent independently of the absolute noise level in the reduced shear. In the following, we identify three principal consequences of this finding. 

First, the noise can be set to a level low enough that (i) the resulting mass distributions can be well approximated with Normal distributions and (ii) individual mass uncertainties (translated into individual bias uncertainties using the true masses) are much narrower than the distribution of all biases. We can thus forgo the somewhat complicated recipe, outlined in \sect~\ref{sec:meth:massbias}, of fitting for the bias distribution, and instead look at the latter directly to understand its properties (such as whether it is, in fact, well approximated by a log-normal). For log-normal distributions, we can then directly find the parameters $\mu$ and $\sigma$. Recalling that $\mu$ is defined as the expectancy value of the distribution of $\bias = \wlmass / \truemass$, we take $\mu = \exp(\overline{\ln \bias})$, while we take $\sigma$ as the standard deviation of $\ln \bias$. Although expected, we have verified explicitly that this yields the same values of $\mu$ and $\sigma$ as the explicit fitting method as the relative noise level approaches zero. Alternative distributions, involving more parameters, can in principle be modeled as linear combinations of log-normal distributions (unless negative masses must be taken into account, as can be the case when including broad miscentring distributions, as seen in \sect~\ref{sec:res:miscenter}), and can therefore by our argument be modeled without adding noise to the reduced shears.

Second, choosing a low enough absolute noise level (or in practice setting the noise to zero and using weights to mimic relative differences of uncertainty between radial bins of reduced shear), the uncertainties in $\mu$ and $\sigma$ depend only on the number of halos in a mass and/or redshift bin. This fact allows us to estimate the number of halos needed to reach a certain level of statistical uncertainty, which propagates as a systematic uncertainty in the subsequent determination of a mass calibration from observations. In particular, we consider an ensemble of $n$ halos in a mass bin at a chosen redshift, from which we estimate $\mu$ and $\sigma$ as $\hat{\mu}$ and $\hat{\sigma}$. Let $\delta_{\ln \mu}$ and $\delta_{\sigma}$ denote the uncertainties in the estimators. These are given by 
\begin{equation}
    \delta_{\ln \mu} = \frac{\sigma}{\sqrt{n}} \approx \frac{\hat{\sigma}}{\sqrt{n}}
\end{equation}
and
\begin{equation}
    \delta_{\sigma} \approx \sigma \sqrt{\frac{1}{2(n-1)}} \approx \frac{\hat{\sigma}}{\sqrt{2n}},
\end{equation}
where the approximations are valid for reasonable estimates of $\sigma$ and sufficiently large $n$ \citep[e.g.][]{EvanHastPeac93}.
With a typical estimate of $\sigma$ around $0.25$ and with $n=100$, $\delta_{\ln \mu} \approx 0.025$, and because $\exp(x) \approx 1+x$ for small $|x|$, this translates into a relative uncertainty in $\mu$ of around 2.5\%. Reaching an accuracy of $1\%$ is thus possible only with hundreds of simulated halos. This poses a challenge, especially in the context of high mass and hydrodynamical simulations with high resolution.  

Third, because the bias of each halo can be estimated independently, it is not strictly necessary to bin the simulation data by redshift and/or mass; instead we can fit some function (e.g. a power law, if applicable) to the full set of data and estimate a functional form for the mass and redshift dependence of the bias distribution.

\subsection{Convergence centers}
\label{sec:disc:kappa}

As noted in \sect~\ref{sec:res:miscenter}, using the convergence SNR peak for centering results in a large positive bias at high noise. This is indeed expected, since we are centering the reduced shear profile on a positive noise peak of the reconstructed convergence, which itself is computed from the noisy reduced shear and thus not independent of it. The effect is less pronounced at high mass, as the SNR improves with a higher signal at constant noise. As the mean bias increases with increasing noise, the bias scatter decreases. This is explained by the fact that the scatter is defined in the space of $\ln \bias$, where $\bias$ is the bias distribution.

\citetalias{2011ApJ...740...25B} found that halo centering errors can introduce negative mass bias at around 5\% when using halo centers from convergence peaks, but did not consider noisy reconstructions of the convergence. Indeed, we find a small negative mean bias when using convergence SNR peaks computed from noiseless reduced shear fields, but with increasing noise the bias quickly becomes positive and reaches levels of up to $50\%$ for weak lensing observations comparable to our setup unless extremely massive clusters are studied.
Considering the low but wide tails of the derived miscentring distributions for convergence centers, we infer that this method is not viable at convergence peaks below a peak signal-to-noise ratio of 4. We have not investigated whether the use of an integrated signal-to-noise measure might improve the situation; such an endeavor is made difficult by the high correlation between pixels in the reconstructed convergence image after Wiener-filtering.   

In deriving the convergence-based miscentring distributions, we defined a search radius of two arcminutes around the most bound particle in the simulation. Analysing actual observations with a similar approach, defining the search radius in terms of the peak in the SZE Comptonization or the X-ray emissivity, could of course lead to the identification of a local convergence peak farther from the unknown bottom of the gravitational well. We performed a rudimentary test for this effect by repeating the analysis of \sect~\ref{sec:res:kappamis:distr}, simulating SZE centers using the miscentring distribution described in \sect~{\ref{sec:res:miscenter}}. For a peak signal-to-noise ratio greater than 4, we found no measurable difference in the resulting miscentring distributions. 

In \sect~\ref{sec:res:kappamis} we derived the SNR-dependent distributions using a constant number density and a constant level of shape noise. Thus, the differences in SNR largely come about because of differences in mass, albeit with considerable scatter due to differences in morphology. As a cross-check, we also tested setups with half or double the noise level and repeated the analysis in terms of miscentring distributions in bins of SNR. Again, we found no discernible effects in the shape and peak positions of the resulting distributions, suggesting that our approach is applicable for different noise levels. The convergence peak centers show by far the strongest dependence on the noise level. Hence, a very careful matching of the noise and mass properties is needed between the real data and simulations if this approach is to be used.

\subsection{Consequences for weak lensing analyses}

There is no straightforward recipe for easily estimating the mass bias for a given weak lensing observation of a given galaxy cluster or sample of galaxy clusters. As we have shown, the bias is indeed critically dependent on the concentration$-$mass relation chosen, as well as on mass, redshift and the radial range in which the mass is constrained from the reduced shear field. Within certain limits, however, it is possible to choose the data analysis in such a way as to minimize these dependencies so that the determination of the bias can be done in fewer mass bins or modeled only at the extremes of the redshifts under consideration. Among the concentration$-$mass relations we have investigated in this paper, it is clear that the model of \cite{2015ApJ...799..108D} (with the corrected parameters set of \citealt{2019ApJ...871..168D}) produces a relatively weak mass dependence. Perhaps surprisingly, the same is the case when using a constant concentration of $c=4$ over the considered mass range. The inner fit radius, $r_\mathrm{min}$, also plays an important role; intermediate values around $0.5$ Mpc produce only moderate mass dependence (with respect to the mass bias mean), while both higher and lower values introduce stronger dependencies up to several percent. 

We have seen that the log-normal distribution is not necessarily a good approximation to the mass bias distribution if significant miscentring is present.
In this case the mass bias distribution cannot be modelled using noiseless simulations in a straightforward way. While a non-parametric model of the bias distribution is possible in principle, it cannot be ruled out that an additional dependence on the noise level would need to be taken into account when additionally accounting for miscentring. 
Alternatively, it is possible to include miscentring in the shear profile model prediction (e.g. \citealt{2012ApJ...757....2G}). 
This is complicated by the fact that the actual amount of miscentring is often poorly constrained for an individual cluster. 
However, a useful approach to approximately accounting for the net impact of miscentring
on a cluster population has recently been described by \cite{2021arXiv210316212G}.

\section{Summary and conclusions}
\label{sec:conclusion}

We summarize our methods and main results as follows:

\begin{enumerate}

\item We use n-body simulations to study the weak lensing mass bias using an azimuthally symmetric model for the reduced shear. While in this work we make use of the NFW model, our methods can be adapted to any radial mass model. Further, we assume a concentration$-$mass relation for our analysis. 

\item To model the effects of miscentring, we adopt various models for miscentring distributions, including X-ray and SZE offset distributions based on hydrodynamical simulations. From the DMO simulations, we also derive miscentring distributions between the peak in the convergence reconstruction and the 3D halo center, which depend on the noise level in the reduced shear measurements.
    
\item Under the assumption of a log-normal distribution of the weak lensing mass bias, we use a bayesian framework for estimating the two parameters of the bias distribution in the presence of shape noise. 
    
\item An important result of this work is the empirical observation that in the presence of an underlying bias distribution that is in fact log-normal, this distribution can be accurately determined from simulations without the need to add shape noise in the analysis. This simplifies the problem of computing the distribution, and makes its determination possible with fewer simulated halos. 

\item We find that in the presence of miscentring, the bias distribution is not log-normal. In particular, the dislocation in the estimated center 
may lead to negative mass estimates even in the absence of shape noise. The resulting bias distribution cannot be captured by the log-normal model. We propose that, given a suitable miscentring distribution, the bias problem be separated from the miscentring problem. Finding accurate miscentring distributions for various observing strategies will be an important task for upcoming large surveys of galaxy clusters, which require highly accurate weak lensing mass estimates.

\item The weak lensing mass bias is dependent on mass and redshift, but also upon observational parameters such as the inner and outer radii of the reduced shear profile. In addition, the bias will vary with the choice of mass density model, including the choice of a concentration$-$mass relation. For this reason, the most viable solution may still be to model the bias of each halo in a sample individually.

\end{enumerate}


\section*{Acknowledgments}

We are grateful to Patrick Simon for making available the Wiener filter implementation
of the convergence reconstruction approach from
\citet{2009MNRAS.399...48S}, and to Matthew Becker for providing the simulation data from \citetalias{2011ApJ...740...25B}. We would further like to thank Peter Schneider for detailed comments that helped to improve the manuscript.

The Bonn group acknowledges support from  the German Federal Ministry for
Economic Affairs and Energy (BMWi) provided through DLR under projects 50OR1803, 50OR2002, 50QE2002, 50OR1407 and 50OR1610 as well as support provided  by the Deutsche Forschungsgemeinschaft (DFG, German Research Foundation) under grant 415537506.

\section*{Data availability}

The data underlying this article will be shared on reasonable request to the corresponding author.


\bibliographystyle{mnras}
\bibliography{martin} 







\bsp	
\label{lastpage}
\end{document}